\documentclass[journal,twoside]{IEEEtran}
\usepackage{times,amsmath}
\usepackage{amsmath,amssymb}
\usepackage{graphicx}
\usepackage{caption}
\usepackage{subcaption}
\usepackage{microtype}
\usepackage{color}
\usepackage[font = footnotesize]{caption}
\usepackage{relsize}

\usepackage{cite}
\usepackage{hyperref}
\usepackage{lettrine}
\usepackage{soul}

\usepackage{etoolbox}
\makeatletter 
\pretocmd\@bibitem{\color{black}\csname keycolor#1\endcsname}{}{\fail}
\newcommand\citecolor[1]{\@namedef{keycolor#1}{\color{blue}}}
\makeatother

\begin{document}

\title{Full-Duplex OFDM Radar\\ With LTE and 5G NR Waveforms:\\ Challenges, Solutions, and Measurements}

\author{Carlos Baquero Barneto,~\IEEEmembership{Student Member,~IEEE,}
        Taneli Riihonen,~\IEEEmembership{Member,~IEEE,}
        Matias Turunen,\\ 
        Lauri Anttila,~\IEEEmembership{Member,~IEEE,}
        Marko Fleischer, 
        Kari Stadius,~\IEEEmembership{Member,~IEEE,}\\
        Jussi Ryyn{\"a}nen,~\IEEEmembership{Senior Member,~IEEE,} and
        Mikko Valkama,~\IEEEmembership{Senior Member,~IEEE}
\thanks{
This paper is an expanded version from the IEEE MTT-S Radio and Wireless Week (Radio and Wireless Symposium, RWS), Orlando, FL, USA, January 20-23, 2019 [10].}
\thanks{This work was partially supported by the Academy of Finland (grants \#288670, \#301820, \#310991, and \#315858), Nokia Bell Labs, and the Doctoral School of Tampere University. The work was also supported by the Finnish Funding Agency for Innovation under the ``RF Convergence'' project.}
\thanks{C. Baquero Barneto, T. Riihonen, M. Turunen, L. Anttila and M. Valkama are with the Unit of Electrical Engineering, Tampere University, Finland.}
\thanks{M. Fleischer is with Nokia Mobile Networks, Germany.}
\thanks{K. Stadius and J. Ryyn{\"a}nen are with Aalto University, School of Electrical Engineering, Finland.}
}


\maketitle
\begin{abstract}
This paper studies the processing principles, implementation challenges, and performance of OFDM-based radars, with particular focus on the fourth-generation Long-Term Evolution (LTE) and fifth-generation (5G) New Radio (NR) mobile networks' base stations and their utilization for radar/sensing purposes. First, we address the problem stemming from the unused subcarriers within the LTE and NR transmit signal passbands, and their impact on frequency-domain radar processing. Particularly, we formulate and adopt a computationally efficient interpolation approach to mitigate the effects of such empty subcarriers in the radar processing. We evaluate the target detection and the corresponding range and velocity estimation performance through computer simulations, and show that high-quality target detection as well as high-precision range and velocity estimation can be achieved. Especially 5G NR waveforms, through their impressive channel bandwidths and configurable subcarrier spacing, are shown to provide very good radar/sensing performance. Then, a fundamental implementation challenge of transmitter--receiver (TX--RX) isolation in OFDM radars is addressed, with specific emphasis on shared-antenna cases, where the TX--RX isolation challenges are the largest. It is confirmed that from the OFDM radar processing perspective, limited TX--RX isolation is primarily a concern in detection of static targets while moving targets are inherently more robust to transmitter self-interference. Properly tailored analog/RF and digital self-interference cancellation solutions for OFDM radars are also described and implemented, and shown through RF measurements to be key technical ingredients for practical deployments, particularly from static and slowly moving targets' point of view.
\end{abstract}

\vspace{-4mm}

\IEEEpeerreviewmaketitle


\section{Introduction}
\label{sec:intro}

\lettrine[lines=2]{\textbf{M}} \enskip ANY EMERGING technologies and applications demand higher and higher communications capacity and bandwidth while the importance of various radio-based sensing schemes is also continuously increasing in commercial, industrial and military fields\cite{7782415,iet_review}. Good examples, where both radio communications and sensing are of high importance, are autonomous cars, or the automotive sector overall, flight control systems, as well as medical sensors and monitoring \cite{5776640,braun2014ofdm,5760698}, to name but a few. While classically radio communications and radio-based sensing systems are designed, developed and deployed completely independently of each other, the congestion of the available radio spectrum has started to raise interest in merging these functionalities and systems to shared frequency bands and potentially even to same hardware platforms. This is commonly referred to as RF convergence in the literature, with comprehensive state-of-the-art surveys being available in \cite{7782415,iet_review}. 

The integrated or joint operation of communications and radar systems implies that the same waveform is utilized for both tasks. It is generally well known that orthogonal frequency division multiplexing (OFDM) waveforms are very well suited for radio communications, providing robustness against multipath fading, facilitating adaptive modulation and coding across subcarriers, as well as offering high flexibility in radio system design and radio resource management. Additionally, the use of OFDM-based waveforms for sensing/radar purposes is receiving increasing interest as described, e.g., in \cite{7782415,5776640, braun2014ofdm,5073387,8053909,8372630,6875584,RWS2019,8647502, 8645165}.
Relatively recently, in  \cite{6875584,8546516,8634255}, a certain level of convergence of mobile communication networks and radar systems was addressed, with a specific focus on the use of the LTE network downlink reference and synchronization signals for passive radar purposes. Additionally, in \cite{8439272}, the use of 5G-grade phased-array hardware for radar purposes is investigated, however, in the context of a classical frequency-modulated continuous wave (FMCW) waveform.

\begin{figure*}[h]
    \centering
    \includegraphics[width=1\textwidth]{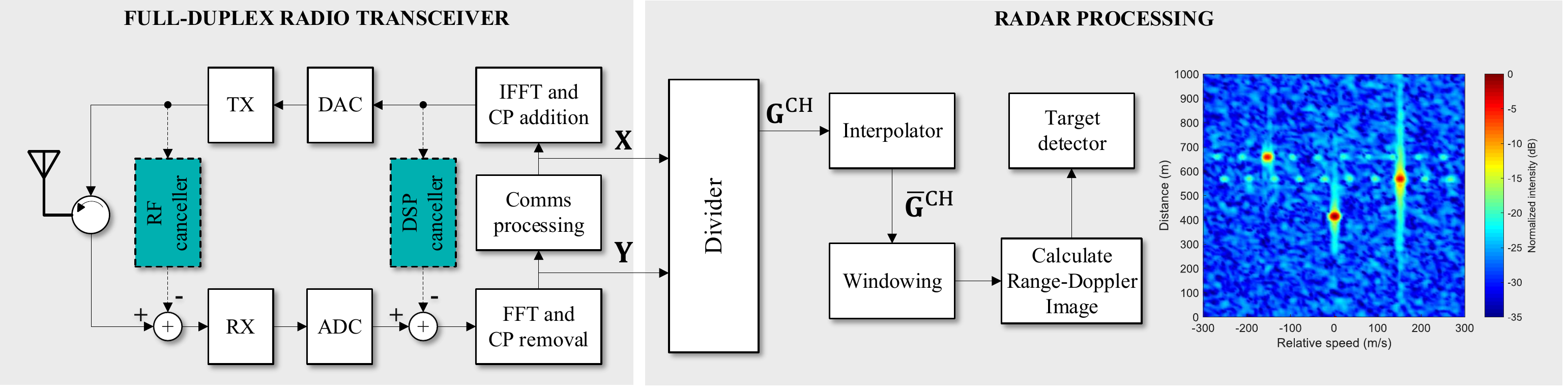}
    \caption{\quad Block diagram of the considered OFDM radar building on regular LTE or NR downlink transmission and frequency-domain radar processing. All communication-related digital waveform processing takes place inside the `Comms processing' box. 
    }
    \vspace{-1mm}
    \label{fig:fig_diagram}
\end{figure*}

In this paper, building on our initial work in \cite{RWS2019}, we also focus on mobile networks, particularly time-division duplexing (TDD)-based LTE and 5G New Radio (NR) systems \cite{2018Dahlman5G}, and the use of the corresponding downlink transmit signals for radar/sensing purposes. We first assume that there is sufficient TX--RX isolation in LTE and NR base stations, referred to as eNB and gNB, respectively, such that when viewed from the sensing point of view, they can essentially act as monostatic radars and thus utilize the complete downlink waveforms for radar processing. We then deploy frequency-domain radar processing, similar to \cite{braun2014ofdm}, while putting specific emphasis on the exact frequency-domain structure of the LTE and NR downlink signals over multiple sub-frames, potentially up to a complete radio frame of 10~ms \cite{3GPPTS36104}, \cite{3GPPTS38104}. We specifically address the impact of null/unused subcarriers within the transmit signal passband of various OFDM symbols in the LTE and NR radio frames and develop a time--frequency interpolation type of an approach to reduce their effects in the radar processing.

Then, we address a fundamental implementation challenge stemming from the limited TX--RX isolation \cite{8642523, 8611240, 8647502, 6832464, 6702851} in real base stations and the corresponding transmitter self-interference at the radar. Our main emphasis is on the challenging monostatic scenario where the same antenna system of the eNB/gNB is shared between the TX and RX, building, e.g., on a circulator \cite{8611240,8335770,8662467,7870377} or an electrical balance duplexer (EBD) \cite{8557219,8645165,8000675}. We first observe that from the perspective of OFDM radar processing, the transmitter self-interference is the most problematic for static target detection while moving targets are already more robust. Then, to enhance the limited TX--RX isolation, properly tailored RF and digital cancellation solutions are devised to suppress the self-interference, without suppressing the actual target reflections. Practical RF measurement results are then also reported, incorporating the developed cancellation solutions, demonstrating the proper operation as well as the target detection and range--velocity estimation capabilities of the overall LTE/NR radar system for both static and moving targets.

Overall, compared to our initial work in \cite{RWS2019} and other existing literature, we have substantially extended the technical scope and contributions by taking the fundamental TX--RX isolation and associated self-interference (SI) issue in monostatic OFDM radars into account. We also provide technical solutions and corresponding RF measurement results to relax or solve the SI challenge, while also incorporate the timely 5G NR aspects at waveform and physical layer processing level. 

For clarity, it is acknowledged that TX--RX isolation is basically a known challenge in all continuous-wave radar systems, and that sufficient RF isolation is always needed to prevent receiver saturation \cite{richards2010principles, 8557219, iet_review, 8645165}. From the actual radar processing point of view, however, FMCW-type of constant-envelope radars \cite{163718,7180336,6810197} can largely suppress the remaining SI through down-mixing with the modulated transmit waveform in the receive path \cite{163718,7180336}. In monostatic OFDM radars, such an approach is not possible but actual cancellation processing is required. We also note that the time-domain digital cancellation solutions described and experimented in this article are complementary methods, compared to radar-domain direct-interference and clutter suppression techniques described, e.g., in \cite{6020778,8645413,8674782}.

Finally, we also acknowledge that an alternative approach to facilitate sensing/radar capabilities in base stations is to deploy a separate radar transceiver system, e.g., FMCW-based, with appropriate coexistence measures with the communications transceiver. However, in this article, we do not pursue such direction but focus on the use of the LTE/NR base station itself for radar purposes, while sending the regular standard-compliant downlink communications waveform, and on the related signal processing aspects and the TX--RX isolation challenge, with particular emphasis on the scenario where the same antenna system is shared by the TX and RX.

The rest of the paper is organized as follows: In Section~\ref{sec:systemModel}, the considered OFDM radar system model is described, incorporating the frequency-domain radar processing based on 3GPP LTE and NR specifications-compliant transmit waveforms. Additionally, the achievable target detection and the associated range and velocity estimation performances are assessed in an ideal self-interference-free reference scenario. 
Then, in Section \ref{sec:SI-canc}, the limited TX--RX isolation and the associated self-interference challenge is addressed, and proper cancellation solutions are developed.
Section \ref{sec:implementation} provides the description of the RF measurement system, obtained results and their analysis.
Finally, Section \ref{sec:conclusion} concludes the work.


\section{OFDM Radar: \\System Model and Reference Performance}
\label{sec:systemModel}

\subsection{System Model and Basic Subcarrier Processing}
The radar functionality is pursued in an LTE or NR network base-station unit by utilizing the known transmit waveform together with frequency-domain radar processing as shown in Fig. \ref{fig:fig_diagram}. The radar processing seeks to detect targets surrounding the eNB/gNB by using the known samples within the LTE or NR frequency-domain resource grid over multiple OFDM symbols, denoted by $\textbf{X}$.
This grid contains frequency-domain samples corresponding to the overall composite transmit waveform, thus comprising all the downlink physical and logical channels and reference signals as specified in \cite{3GPPTS36104} for LTE and in \cite{3GPPTS38104} for NR. The size of the grid or matrix $\textbf{X}$ is $S\times R$ where $S$ denotes the number of active subcarriers while $R$ indicates the number of OFDM symbols considered in the radar processing.

In the transmitter processing, the time-domain waveform is generated through block-wise IFFT operating on $\textbf{X}$, together with cyclic prefix addition, as described in \cite{3GPPTS36104}, \cite{3GPPTS38104}. The radiated transmit waveform will then interact with one or multiple targets, producing reflections that will be captured by the receiving system. For simplicity, perfect isolation between the transmit and receive systems is assumed in this section, while the limited TX--RX isolation aspects and the associated self-interference challenge are addressed in Section III. The received signal containing the target reflections is demodulated and processed through FFT to obtain the corresponding receive grid, denoted by $\textbf{Y}$.

The distance and the relative speed of a single target correspond to a propagation delay $\tau_{k}$ and a Doppler shift $f_{D,k}$, respectively. 
With $K$ point targets or reflections, the receive grid sample on $p$th row and $q$th column, i.e., $Y_{p,q}=(\textbf{Y})_{p,q}$ can be expressed as \cite{RWS2019,braun2014ofdm} 
\begin{equation}
\label{eq_Ypq}
    Y_{p,q} = \sum_{k=0}^{K-1} b_{k} 
    X_{p,q} 
    e^{2\pi j(q T_{s}f_{D,k}-p\tau_{k}\Delta f)} + N_{p,q}
\end{equation}
where $X_{p,q}=(\textbf{X})_{p,q}$, $N_{p,q}$ corresponds to receiver thermal noise sample, while $b_{k}$ models the effective attenuation factor of the $k$th reflection. Furthermore, the subcarrier spacing and the total OFDM symbol duration (including the cyclic prefix of length $T_{cp}$) are denoted by $\Delta f$ and $T_{s}=1/\Delta f + T_{cp}$, respectively. In general, the propagation delay causes different phase shifts for different subcarriers which can be used to estimate the range of each target. Similarly, the Doppler shift produces different phase shifts for the different OFDM symbols that can be utilized to compute the relative speeds of the targets.

The actual radar processing building on the known transmit waveform can be implemented in multiple ways \cite{Levanon2004radarsignals,richards2010principles}, either with time-domain or subcarrier-domain processing. In this work, similar to, e.g., \cite{5776640,braun2014ofdm,5494616,5760698,5073387,8647502}, we adopt subcarrier-domain processing utilizing directly the transmit and receive grids of samples, i.e., $\textbf{X}$ and $\textbf{Y}$, respectively. For the classical matched filtering-based approach \cite{richards2010principles}, represented at OFDM symbol level, a new processed matrix $\textbf{G}$ can be defined as $\textbf{G}^\mathrm{MF} = \textbf{Y} \odot \textbf{X}^*$,
where $\odot$ refers to element-wise (Hadamard) product while the superscript $(\cdot)^*$ denotes complex conjugation. Alternatively, the reflection channel estimation-like scheme, utilized also in \cite{5776640,braun2014ofdm,8647502,5073387}, reads $\textbf{G}^\mathrm{CH} = \textbf{Y} \oslash \textbf{X}$, 
where $\oslash$ denotes element-wise division.

Based on (\ref{eq_Ypq}), when interpreted at an individual subcarrier $p$ of an OFDM symbol $q$, we can write $G^\mathrm{MF}_{p,q}=(\textbf{G}^\mathrm{MF})_{p,q}$ as
\begin{equation}
\begin{split}
\label{eq_G_MF_pq}
    G^\mathrm{MF}_{p,q} &= Y_{p,q}X^*_{p,q} \\
    &= \sum_{k=0}^{K-1} b_{k} |X_{p,q}|^2 e^{2\pi j(q T_{s}f_{D,k}-p\tau_{k}\Delta f)} + \tilde{N}^{\mathrm{MF}}_{p,q}
\end{split}
\end{equation}
while the corresponding expression for $G^\mathrm{CH}_{p,q}=(\textbf{G}^\mathrm{CH})_{p,q}$ reads
\begin{equation}
\begin{split}
\label{eq_G_CH_pq}
    G^\mathrm{CH}_{p,q} &= \frac{Y_{p,q}}{X_{p,q}} \\
    &= \sum_{k=0}^{K-1} b_{k} e^{2\pi j(q T_{s}f_{D,k}-p\tau_{k}\Delta f)} + \tilde{N}^{\mathrm{CH}}_{p,q}
\end{split}
\end{equation}
where $\tilde{N}^{\mathrm{MF}}_{p,q}$ and $\tilde{N}^{\mathrm{CH}}_{p,q}$ denote processed noise samples. 
While the matched filter is known \cite{Levanon2004radarsignals,richards2010principles} to optimize the signal-to-noise ratio (SNR) from an individual target or reflection point of view, the noiseless range and velocity profiles stemming from (\ref{eq_G_MF_pq}) are clearly data-dependent. Instead, the noiseless component in (\ref{eq_G_CH_pq}) building on the channel estimation approach is data-independent. Additionally, as discussed in \cite{5073387,5776640}, correlator or matched filter-based processing can have some drawbacks related to range profile sidelobes and thus potentially more limited dynamic range in scenarios with multiple targets. This is inline with the ambiguity function \cite{Levanon2004radarsignals} based analysis carried out in \cite{8634255,6875584} for OFDM signals. Hence, in the rest of this article, similar to \cite{5776640,braun2014ofdm,8647502,5073387}, we follow the channel estimation-like approach and assume that the subcarrier-level pre-processing is carried out as defined in (\ref{eq_G_CH_pq}).

\subsection{Interpolation, Target Detection and Range--Velocity Estimation}

For fixed $p$ and varying $q$, the samples of $G^\mathrm{CH}_{p,q}$ correspond to a sum of complex exponentials with oscillating frequencies being defined by the Doppler shifts, and thus the target velocities. 
Similarly, if $q$ is fixed while $p$ varies, we again obtain a sum of complex exponentials but now the frequencies are defined by the propagation delays, and thus the target distances. These facilitate both target detection as well as velocity and range estimation, as defined below. 
However, the LTE and NR transmit grids contain unused subcarriers within the transmit signal passband, whose locations also vary from OFDM symbol to another \cite{3GPPTS36104},\cite{3GPPTS38104}. The direct calculation of (\ref{eq_G_CH_pq}) is therefore not feasible in those points, but it is carried out only for the active subcarriers. Then, to deal with the missing samples prior to the actual target detection and range--velocity estimation, the proposed LTE/NR-compliant radar processing includes appropriate interpolation along either $p$ or $q$ domain.

Stemming from the more detailed structure of the LTE and NR grids \cite{3GPPTS36104}, \cite{3GPPTS38104}, we adopt linear interpolation across OFDM symbols, and thus define the interpolated grid sample $\bar{G}^\mathrm{CH}_{p,q}$ at an unused subcarrier $p$ of an OFDM symbol $q$ as
\begin{equation}
\label{interp}
\bar{G}^\mathrm{CH}_{p,q} = {G}^\mathrm{CH}_{p,q_1} + 
\left ( \frac{{G}^\mathrm{CH}_{p,q_2} - {G}^\mathrm{CH}_{p,q_1}}{q_2-q_1}\right )(q-q_1)
\end{equation}
where $q_1<q<q_2$ and the subcarrier $p$ is assumed active at OFDM symbols $q_1,q_2$. In the rest of this article, we use systematically the notation of the interpolated grid, i.e., $\bar{G}^\mathrm{CH}_{p,q}$, while naturally at the active subcarriers $\bar{G}^\mathrm{CH}_{p,q}={G}^\mathrm{CH}_{p,q}$.

In this work, the target detection and range--velocity estimation build on the range--Doppler profile calculated by using the interpolated grid.
One common approach \cite{5494616} is to calculate the radar image through the periodogram, expressed as
\begin{equation}
\label{eq_periodogram}
    A(s,r)=\left |  
        \sum_{q=0}^{R'-1} \left ( \sum_{p=0}^{S'-1}  (\bar{G}^\mathrm{CH}_{p,q} W_{p,q} 
        e^{j2\pi \frac{p\,s}{S'}}  \right )
        e^{-j2\pi \frac{q\,r}{R'}}
    \right |^2
\end{equation}
where the inner IFFTs of size $S'\geq S$ yield OFDM symbol-wise range profiles while the outer FFTs of size $R'\geq R$ correspond to velocity profiles. Importantly, the ranges of $s$ and $r$ over which the transforms are calculated can be limited to chosen feasible distances and velocities, respectively. These are denoted in the following by $S_\mathrm{MAX}$ and $R_\mathrm{MAX}$, while the corresponding set of periodogram indices is denoted by $\Omega_A$.
The periodogram in (\ref{eq_periodogram}) includes also sets of additional weights ${W_{p,q}}$ which correspond to windowing of the interpolated grid, in order to control the sidelobe levels \cite{5494616}.

In the actual target detector, the periodogram is subject to a threshold test \cite{Levanon2004radarsignals,richards2010principles}, expressed as
\begin{equation}
\label{eq_threshold}
    A(s,r) \hspace{2mm} \substack{H_0\\\lessgtr\\H_1} \hspace{2mm} T_\mathrm{th}
\end{equation}
where $(s,r) \in \Omega_A$, $T_\mathrm{th}$ denotes the detection threshold, $H_0$ refers to the null hypothesis (no target, noise only), while $H_1$ refers to the alternative hypothesis (a target present). Specifically, if $A(s,r) > T_\mathrm{th}$, the detector declares that the target is present, while if $A(s,r) < T_\mathrm{th}$, it declares that the target is absent. We then denote the probability that $A(s,r) > T_\mathrm{th}$, under $H_0$ being true, by $P_{\mathrm{FA}}$. The corresponding total false alarm probability in the whole search space $\Omega_A$ then reads $P_{\mathrm{FA,tot}} = 1-(1-P_{\mathrm{FA}})^{|\Omega_A|}$ where $|\Omega_A|$ refers to the size of $\Omega_A$. As shown, e.g., in \cite{braun2014ofdm,richards2010principles}, assuming Gaussian-distributed receiver noise, the detection threshold $T_\mathrm{th}$ can be straightforwardly calculated to fix the $P_{\mathrm{FA,tot}}$ to a chosen value, commonly known as the constant false-alarm rate (CFAR) detector.  

In the case of the threshold test declaring that a target is present, the estimation of the target range and velocity can be pursued \cite{Levanon2004radarsignals,richards2010principles}. Considering, for presentation simplicity, the single-target scenario, it is shown in \cite{5494616} that the location of the peak of the periodogram is the corresponding maximum likelihood estimator, i.e.,
\begin{equation}
    \label{MLest}
    (\hat{s},\hat{r}) = \underset{(s,r) \in \Omega_A}{\mathrm{arg} \hspace{1mm} \mathrm{max}} \hspace{1mm} A(s,r)
\end{equation}
To improve the resolution beyond the basic pixel size of the range--Doppler map, different interpolation-type peak refinement methods can be adopted as discussed, e.g., in \cite{braun2014ofdm}. For generality, we note that the periodogram-based approach in (\ref{eq_periodogram}) is only one feasible approach to create test statistics for target detection, while alternative methods can also be pursued.

\subsection{Reference Performance with LTE/NR Carriers}
\label{sec:simulations}
 
\begin{table}[t!]
  \begin{center}
  \setlength{\tabcolsep}{4.5pt}
    \caption{\textsc{ Basic features and OFDM radar parameters for the \break considered LTE and NR channel bandwidths at 3.5 GHz}}
    \label{tab:simulation_parameters}
    \begin{tabular}{|c || c | c | c | c|}
      \hline
      \textbf{ } & \textbf{LTE 20} &  \textbf{NR 20} &  \textbf{NR 40} & \textbf{NR 100}\\
      \hline 
      \hline
      Passband width (MHz)	& 18 & 19.08 & 38.16 & 98.28\\
      Subcarrier spacing $\Delta f$ (kHz)	& 15 & 15 & 30 & 30\\
      Cyclic prefix length, $T_{cp}$ ($\mu$s) & 4.7 & 4.7 & 2.3 & 2.3 \\
      Active subcarriers, $S$	& 1200 & 1272 & 1272 & 3276\\
      OFDM symbols, $R$	& 140 & 140 & 280 & 280\\
      Bandwidth utilization ($\%$)	& 90 & 95.4 & 95.4 & 98.28\\
      \hline
      Distance resolution (m)	& 8.3 & 7.9 & 3.9 & 1.5\\
      Maximum distance (m)	& 700 & 700 & 350 & 350\\
      Velocity resolution (m/s)	& 4.2 & 4.2 & 4.2 & 4.2\\
      Maximum velocity  (m/s)	& $\pm$64 & $\pm$64 & $\pm$128 & $\pm$128\\
      \hline
    \end{tabular}
  \end{center}
\end{table}
\begin{figure}[t!]
    \centering
    \begin{subfigure}{1\columnwidth}
        \centering
        \includegraphics[width=1\columnwidth]{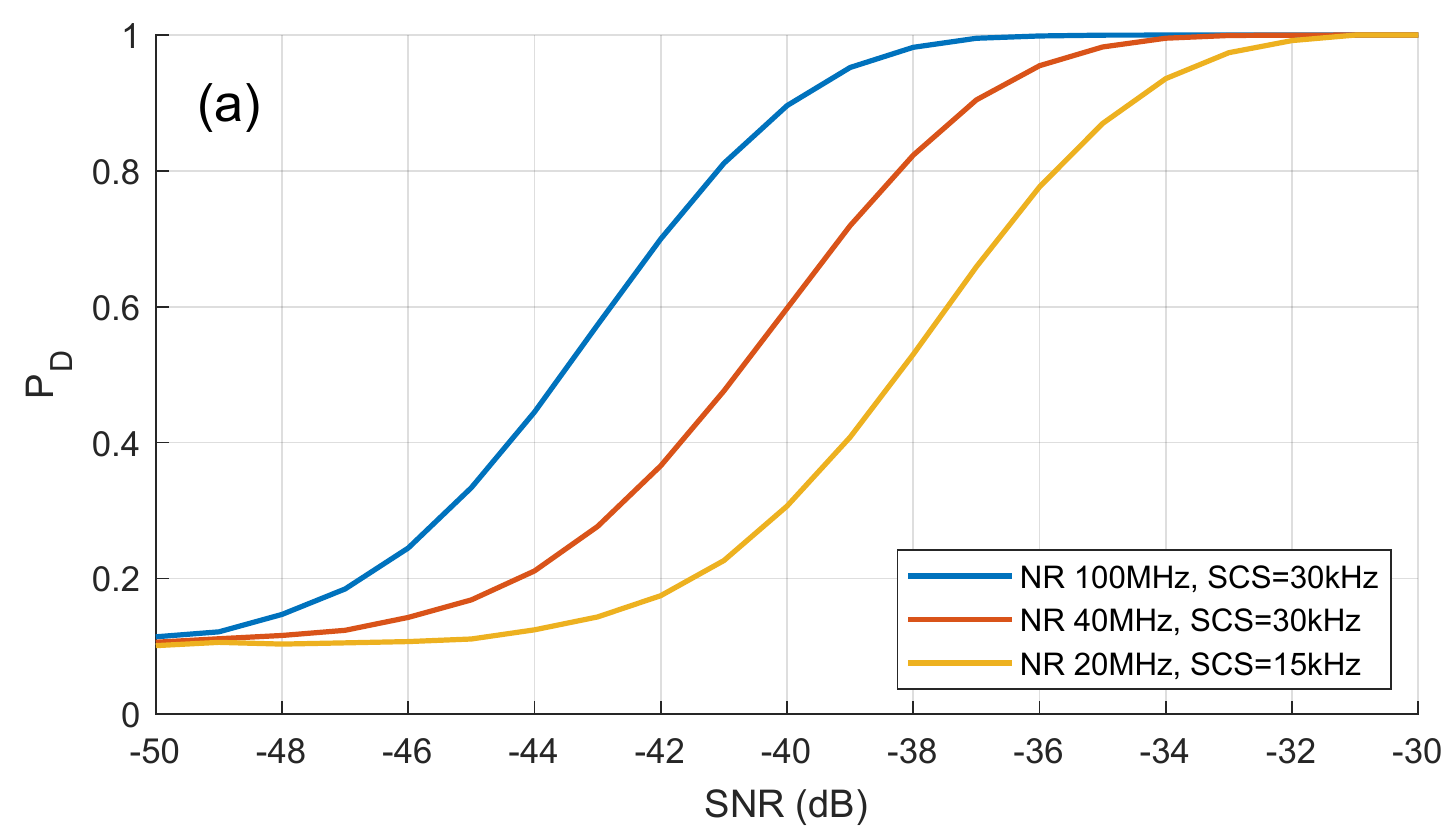}
        \label{fig:LTEvsNR_Pd}
    \end{subfigure}
    \hfill
    \begin{subfigure}{1\columnwidth}
        \centering
        \includegraphics[width=1\columnwidth]{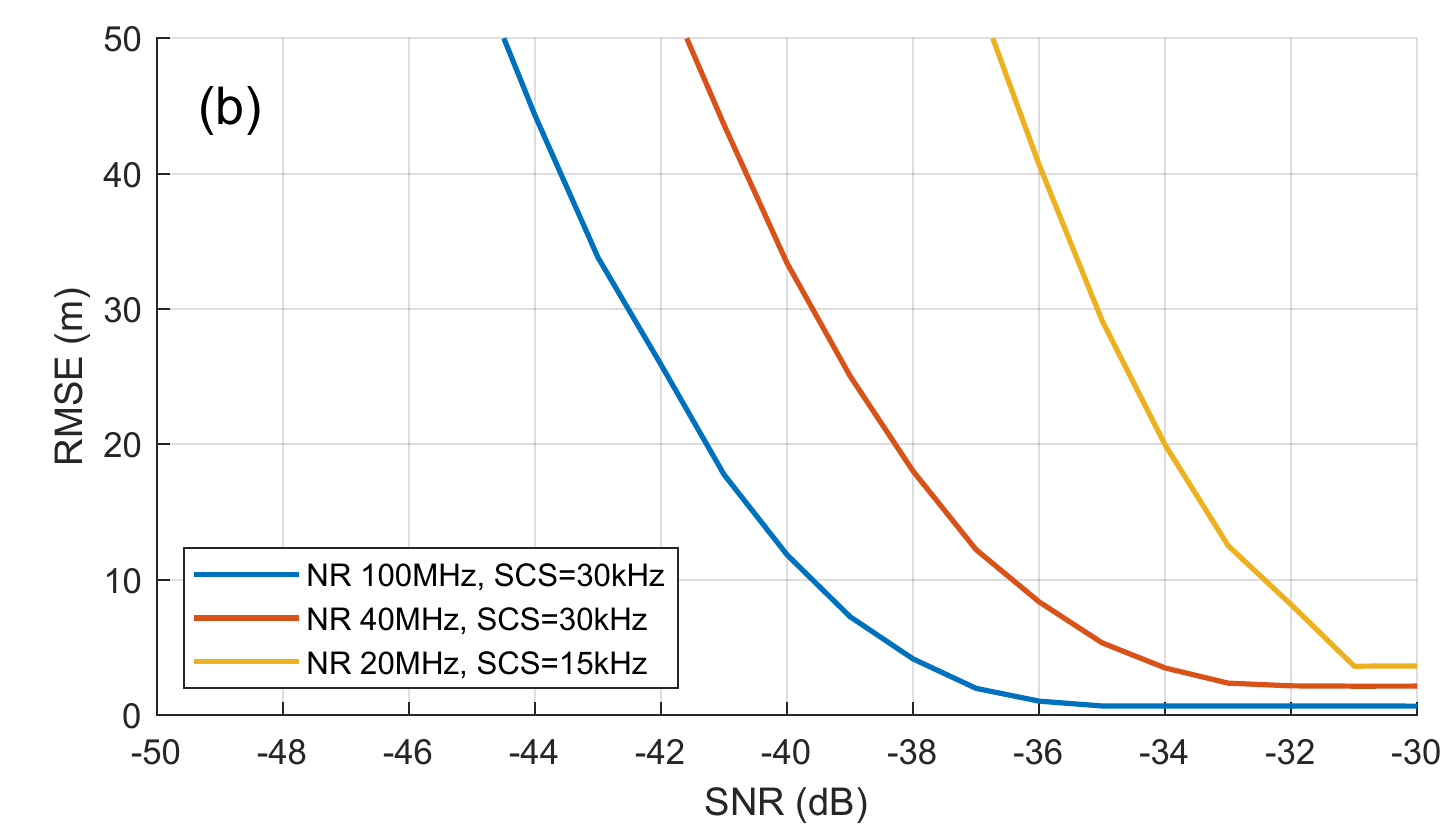}
        \label{fig:LTEvsNR_distance}
    \end{subfigure}
    \hfill
    \begin{subfigure}{1\columnwidth}
        \centering
        \includegraphics[width=1\columnwidth]{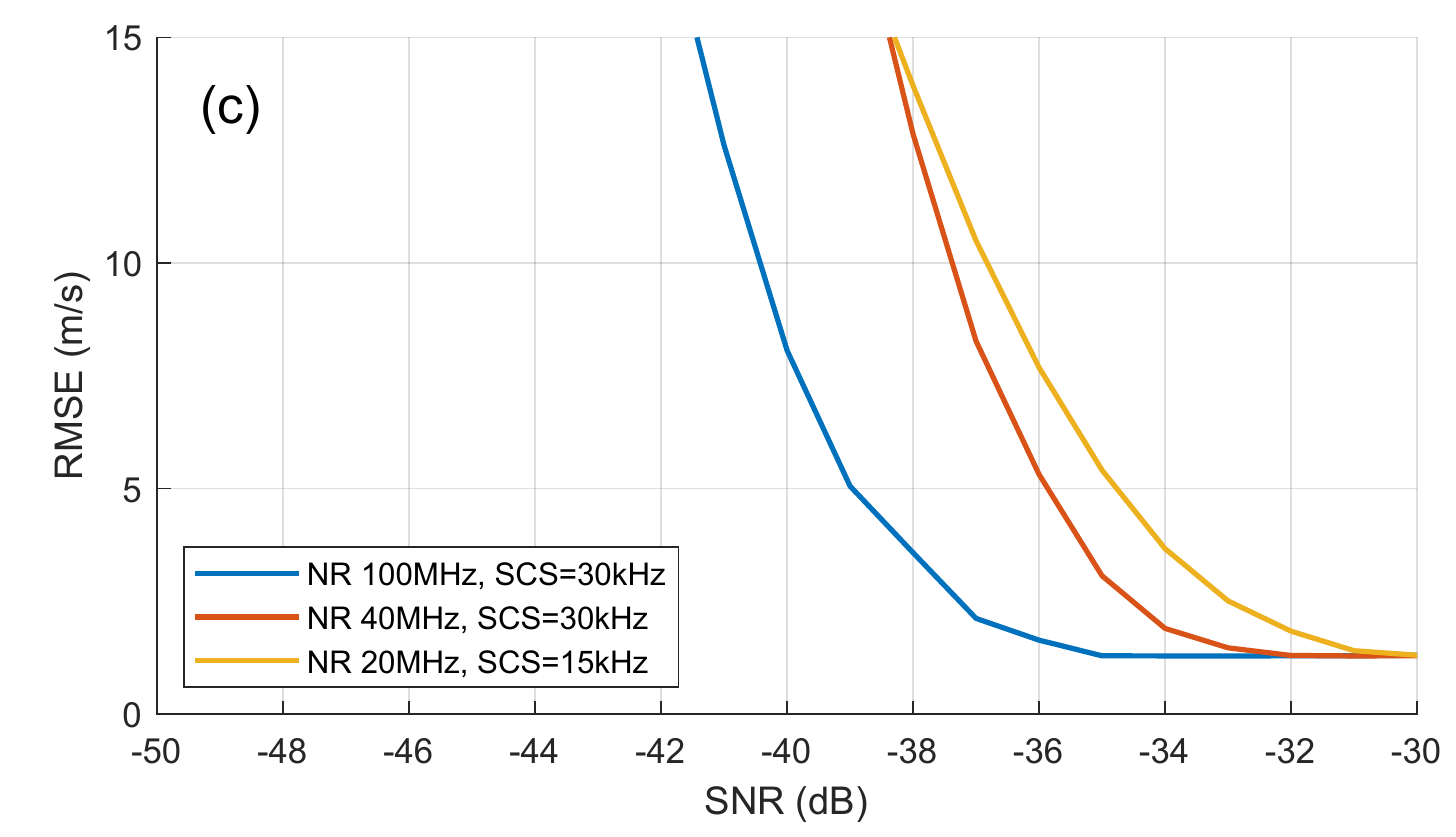}
        \label{fig:LTEvsNR_velocity}
    \end{subfigure}
    \caption{\quad Single-target radar performance at 3.5~GHz as a function of receiver input SNR, in terms of (a) detection probability ($P_D$), (b) target distance estimation RMSE, and (c) target velocity estimation RMSE, for 10~ms NR waveforms with 20~MHz, 40~MHz and 100~MHz carrier bandwidths.}
    \label{fig:LTEvsNR}
\end{figure}

\begin{figure*}[t!]
    \centering
    \begin{subfigure}[t]{0.31\textwidth}
        \centering
        \includegraphics[width=1\columnwidth]{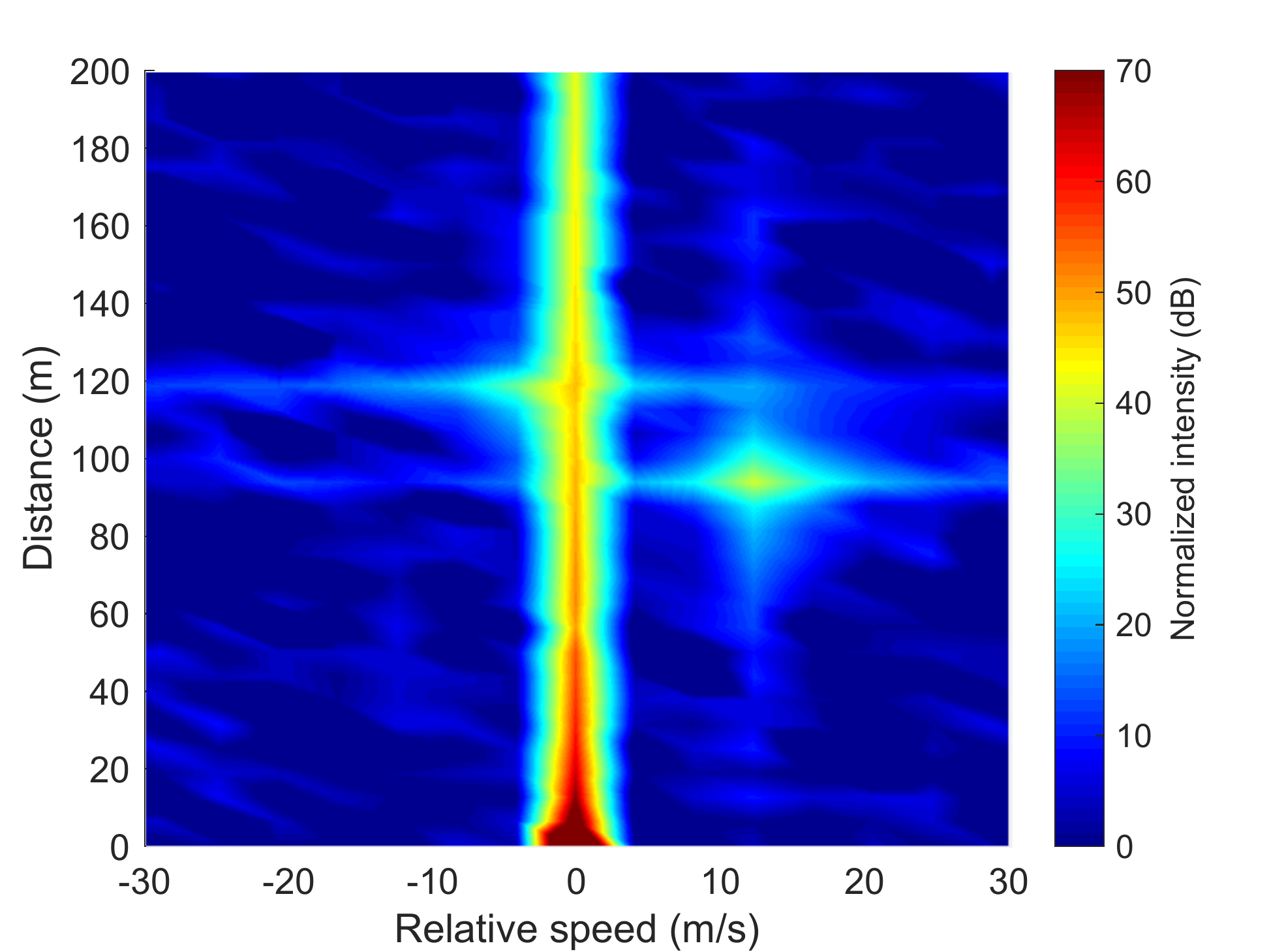}
        \caption{}
        \label{fig:SI_70dB}
    \end{subfigure}
    \begin{subfigure}[t]{0.31\textwidth}
        \centering
        \includegraphics[width=1\columnwidth]{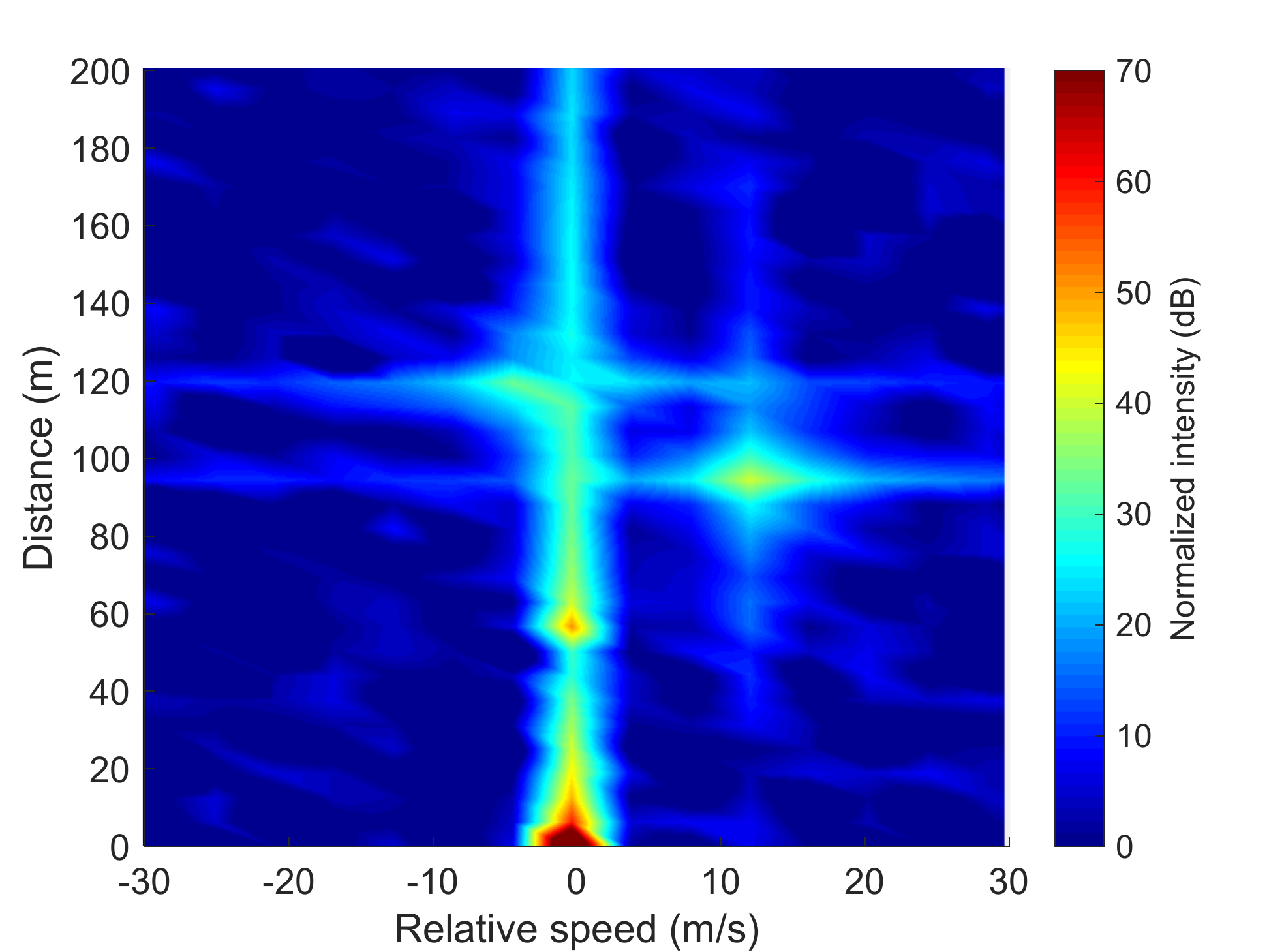}
        \caption{}
        \label{fig:SI_50dB}
    \end{subfigure}
     \centering
    \begin{subfigure}[t]{0.31\textwidth}
        \centering
        \includegraphics[width=1\columnwidth]{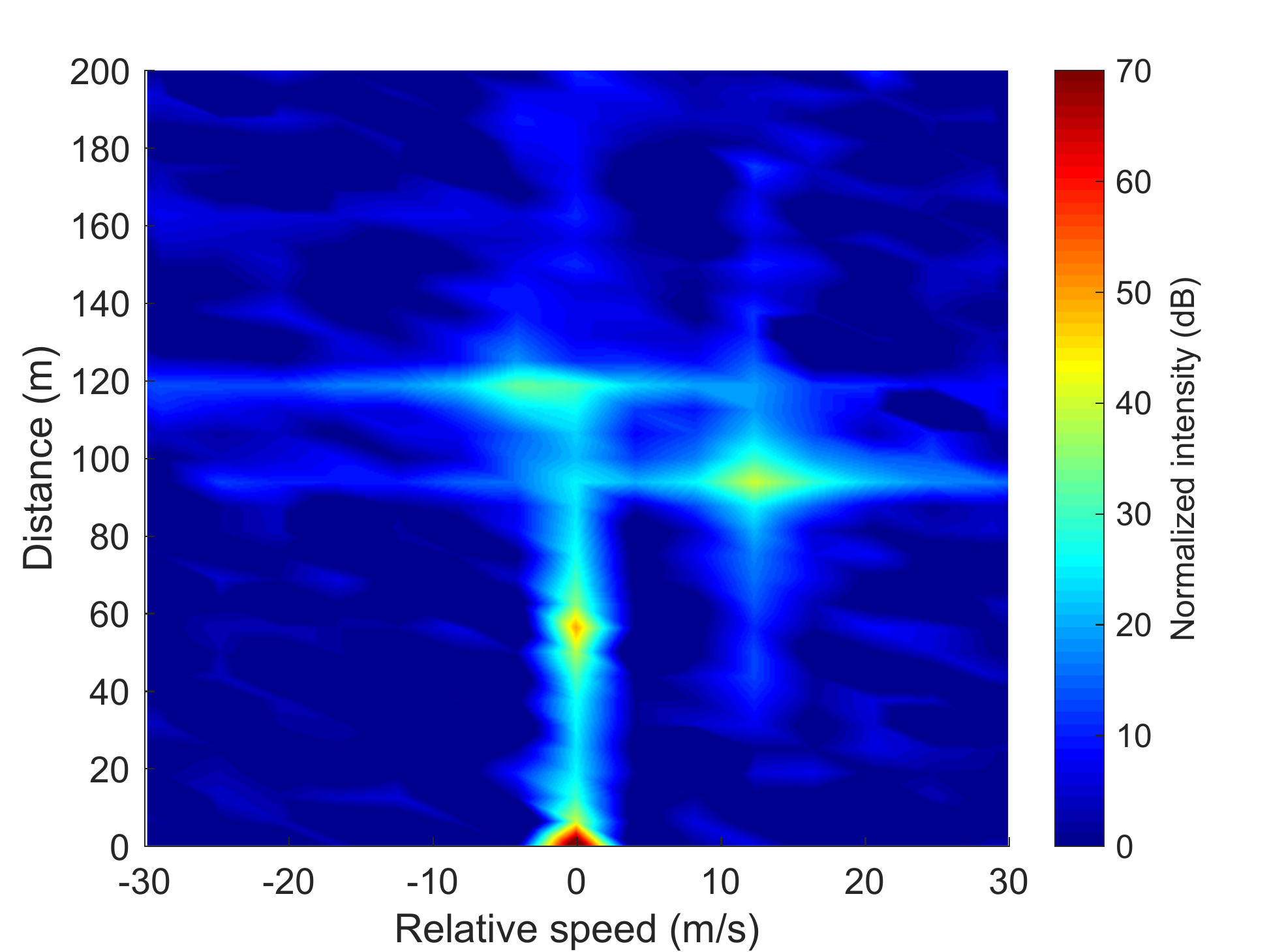}
        \caption{}
        \label{fig:SI_30dB}
    \end{subfigure}
    \vspace{-1.5mm}
    \caption{\quad Example radar images in a scenario with three true targets located at 60~m, 90~m and 120~m distances and moving with relative velocities of 0~m/s, $+$12~m/s and $-$2~m/s. The radar receiver operates under direct SI levels of (a) 70~dB, (b) 50~dB and (c) 30~dB with respect to the thermal noise floor. NR waveform of 10~ms with channel bandwidth of 40~MHz and subcarrier spacing of 30~kHz is used.
 }
  \vspace{-2mm}
    \label{fig:SI_analysis}
\end{figure*}

Simulations are next carried out to evaluate the performance of the interpolated periodogram-based OFDM radar described in the above subsection, and to compare the LTE and NR waveforms and their feasibility for high-precision range and velocity estimation. In the evaluations, LTE and NR waveforms with carrier bandwidth of 20~MHz and subcarrier spacing of 15~kHz are considered as the baseline. Additionally, NR waveforms with 30~kHz subcarrier spacing and carrier bandwidths of 40~MHz and 100~MHz are also adopted. Assuming a downlink transmit radio frame of 10~ms \cite{3GPPTS36104},\cite{3GPPTS38104}, and the above radio interface numerologies, the main OFDM radar parameters are summarized in Table~\ref{tab:simulation_parameters}.

The assumed network center frequency is $f_c=3.5$~GHz, and all the main downlink physical channels and signals, such as PSS, SSS, CRS, DMRS, CSI-RS, PRS, PDSCH, PDCCH, PCFICH, PHICH, PBCH, are modelled in both LTE and NR sides when applicable. For simplicity, $R'$ and $S'$ are set to $R$ and $S$, respectively.
As shown in \cite{braun2014ofdm}, the basic distance and velocity resolutions of the OFDM radar read $\Delta d = c/(2 S \Delta f)$ and $\Delta v = c/(2 R  T_s f_c)$, respectively, with $c$ denoting the speed of light, which correspond to the resolution in the range--Doppler map. The resolution  values for different waveform configurations are also shown in Table~\ref{tab:simulation_parameters}. No further peak refinement methods are adopted in these simulations.
Finally, a basic rectangular window is used as the window function in (\ref{eq_periodogram}). 

In these reference evaluations, only a single target is present, and we vary the true distance and speed of the target randomly, from realization to another, with uniform distributions within 20~to~200~m (distance) and $-$40~to~40~m/s (velocity).
In the radar processing, no prior information about the target distance or speed is assumed, and thus in the periodogram calculations and the corresponding detector, the overall feasible space $\Omega_A$ contains distances up of a one-way propagation delay of half the cyclic prefix, and velocities up to a Doppler shift of $\pm$10\% of the subcarrier spacing. The distance limit is directly stemming from the subcarrier-based processing, to avoid intersymbol interference (ISI), while the velocity/Doppler limit is a practical engineering rule of thumb, to keep intercarrier interference (ICI) tolerable \cite{braun2014ofdm,2018Dahlman5G}. The detection threshold $T_\mathrm{th}$ is set such that $P_{\mathrm{FA,tot}}=10\%$. The numerical values of these limits are also shown in Table~\ref{tab:simulation_parameters}. 

We also consider multipath components, i.e., there are indirect reflected and/or scattered waves entering the receiver in addition to the direct reflection. The total power of the reflected and scattered multipath components is $-15$~dB, relative to the direct reflected component, with an RMS delay spread of 100~ns, representing a realistic example scenario. 

Fig.~\ref{fig:LTEvsNR} presents the obtained probability of target detection for varying receiver input SNR, as well as the corresponding distance and velocity estimation error behavior---calculated only when target detection is positive---for the considered NR channel bandwidth and subcarrier spacing cases shown in Table~I. LTE 20 MHz is eventually not shown since its performance is essentially identical to NR 20 MHz.  
As can be observed, reliable target detection can be achieved down to input SNRs in the order of $-30$~dB to $-40$~dB, 
the exact number depending on the channel bandwidth and the required probability of detection. In general, as discussed in \cite{5073387,5776640}, the subcarrier-based radar processing provides a processing gain of
\begin{equation}
\label{processing_gain}
    \Gamma_{\mathrm P} = 10 \log_{10}(R \times S)
\end{equation}
decibels against thermal noise, thus with given receiver input SNR the larger processing gains obtained through larger bandwidths improve the detection, as shown in Fig.~\ref{fig:LTEvsNR}(a).

Additionally, the results in Fig.~\ref{fig:LTEvsNR} illustrate that at those SNR values, where high detection probabilities are obtained, also the distance and velocity estimation errors are well-behaving. The results also clearly show that when the probablity of detection is approaching 100\%, the estimates' RMSE values are essentially directly defined by the resolution values shown in Table~\ref{tab:simulation_parameters}. In other words, the RMSE values are converging towards $\sqrt{(\Delta d)^2 / 12}$ and $\sqrt{(\Delta v)^2 / 12}$ for distance and velocity, respectively, reflecting uniform error distribution within the pixel width.

Overall, we observe that the large carrier bandwidths available in NR, even at sub-6~GHz frequencies \cite{3GPPTS38104}, build a basis for good target estimation accuracy and precision, particularly when it comes to the distance estimation.
In general, 5G NR networks can be deployed also at 24--40~GHz bands \cite{3GPPTS38104}, with continuous carrier bandwidths up to 400~MHz available already in \mbox{3GPP Release 15}, which together with the larger center frequencies facilitate further improvements in both distance and velocity estimation resolution.


\section{Self-interference Problem and\\Cancellation Solutions}
\label{sec:SI-canc}

\subsection{TX--RX Isolation and Self-interference Problem}
In ordinary TDD-based communication networks, the transmit and receive functionalities are divided in time and thus a base station's transmitter and receiver do not need to operate simultaneously. However, in both LTE and NR networks, the minimum downlink allocation within a radio frame contains seven~OFDM symbols, or one slot, which corresponds to 0.5~ms with 15~kHz subcarrier spacing. Hence, from the point of view of the considered OFDM radar concept, the receiver must be operating simultaneously unlike in a pulsed radar, otherwise no targets within tens of kilometers could be detected rendering the whole concept useless.

The fact that the receiver must be operating simultaneously while transmitting, and at the same carrier frequency, leads to large implementation challenges particularly in facilitating sufficient transmitter--receiver isolation \cite{8645165,richards2010principles,8642523,bharadia2013full,6832471,7565192,8053906,8335770}. This calls for new hardware, to replace the TDD RF switching with more elaborate circulator
 \cite{8611240,8335770,8662467,7870377} or EBD \cite{8557219,8645165,8000675} type of circuitries, and has been studied actively over the recent years, under the inband full-duplex radio terminology, with primarily communications applications in mind, see, e.g., \cite{8642523,bharadia2013full,6832471,7565192,8053906,8335770,7756408,5985554,7105651,6736751,7105647}. In the radar context, the self-interference (SI) stemming from the direct coupling of the transmit signal to the receiver can be interpreted as a strong static target at a very short distance \cite{8645165,8647502,8528501,8645413}. Hence, one could argue that as long as a certain level of TX--RX isolation can be provided, such that the low noise amplifier (LNA) and the rest of the sensitive receiver electronics can tolerate the remaining SI, the true echoes stemming from true targets can be separated in the radar processing. Recent examples of radar-domain processing solutions for direct interference and clutter suppression can be found, e.g., in \cite{6020778,8674782}, developed primarily for passive radar applications, reporting commonly some 15--20~dB processing gain at best. 

However, since the eNB/gNB transmit power can be even more than 140~dB larger than the receiver thermal noise floor, facilitating sufficient TX--RX isolation as a whole is technically very challenging, particularly in the monostatic shared-antenna OFDM radar case with limited passive isolation and high peak-to-average power ratio (PAPR) in the transmit waveform. An extreme example is a macro base station that commonly utilizes a transmit power in the order of +46~dBm, while the receiver noise floor assuming 20~MHz channel bandwidth and 4~dB noise figure is only -97~dBm. For one, with such transmit powers and large PAPR, preventing LNA and receiver saturation is already a fundamental problem. Additionally, the powerful SI component in the receiver digital signal can largely mask the true echoes and targets---particularly those that are static, but also other slowly moving targets---rendering thus very tough requirements for the digital-domain suppression algorithms. 

A concrete example of the latter problem is given in 
Fig.~\ref{fig:SI_analysis}, showing the OFDM radar-based range--velocity profiles, i.e., radar images, when the reflections of three true targets are observed and processed by the receiver, under different levels of the direct SI. In this case, an NR signal with channel bandwidth of 40~MHz and subcarrier spacing of 30~kHz is used, and a complete 10~ms radio frame is transmitted and processed. The direct SI entering the receiver is either 70~dB, 50~dB or 30~dB above the receiver thermal noise floor, corresponding to subfigures (a), (b) and (c), respectively, and contains only a single coupling path with nanosecond scale delay. It can be clearly observed that the direct SI is reflected in the radar processing  as a strong static target located essentially at zero distance. Furthermore, and importantly, the sidelobes of the SI produce a substantial masking effect which largely complicates the detection of true weaker targets moving with lower velocities along a wide range of distances, despite the true SI has negligibly small physical coupling delay and zero Doppler.  

In general, as shown in (\ref{processing_gain}), the subcarrier-based radar processing presented in Section \ref{sec:systemModel} provides large processing gain against noise.
This thus provides a corresponding SNR improvement for detecting weak target reflections way below the thermal noise floor, as demonstrated already along Fig.~\ref{fig:LTEvsNR}. However, similar to clutter, the direct SI component is also subject to the same processing gain, and therefore this does not solve the isolation challenge. Additionally, as the basic power calculations show above, more than 100~dB of total SI suppression is required, which calls for multiple complementary methods as no single technique can facilitate such high isolation.
Hence, we can conclude that on top of basic passive RF isolation and radar-domain digital suppression methods, efficient active RF and digital SI cancellation methods are needed, particularly from the static and slow-moving targets' viewpoint, as well as overall to prevent receiver saturation.

\subsection{RF and Digital Cancellation Solutions}
As identified above, facilitating sufficient TX--RX isolation is a key technical ingredient in OFDM radars. In this work, on top of passive isolation, we primarily focus on active RF cancellation and time-domain digital cancellation methods. Compared to the existing inband full-duplex radio research, a specific radar-related flavor is that only the direct SI should be cancelled or suppressed, along with possible reflections from very close-by surfaces, while the echoes from true targets must be preserved. This is one clear difference to all earlier full-duplex radio works, such as \cite{7146163,7756408,6353396}, that do not separate between the direct and reflected SI components. Additionally, we note that agnostic digital cancellation techniques, independent of the specific radar processing approach are pursued, which can then be complemented with, e.g., the CLEAN-like methods \cite{6020778,8647502,8645413,8645165} operating in the radar domain.

In general, despite a monostatic shared-antenna OFDM radar is pursued, the so-called direct SI can contain frequency selectivity in its coupling response \cite{6353396,7105647}. Hence, a properly devised multitap RF canceller is adopted, seeking to protect the LNA and the following receiver chain. A similar architecture as in our earlier work in \cite{7146163} is adopted such that the RF canceller uses the power amplifier (PA) output as a reference, tapping it through multiple parallel RF delays and vector modulators to obtain an accurate estimate of the direct SI. 

In this work, we consider three RF taps, and formally express the signal after the RF canceller as  $y_{\mathrm{RF}}(t) = r_{\mathrm{RF}}(t) - \hat{s}_{\mathrm{RF}}(t)$ where $r_{\mathrm{RF}}(t)$ is denoting the receiver input signal while the actual RF cancellation signal $\hat{s}_{\mathrm{RF}}(t)$ reads
\begin{equation}
\label{RF_canceller_1}
      \hat{s}_{\mathrm{RF}}(t) = \sum_{l=1}^{3}\alpha_l  \left(\mathrm{cos}(\theta_l) x_{\mathrm{RF}}^{\mathrm{0}}(t-\tau_l) - \mathrm{sin}(\theta_l) x_{\mathrm{RF}}^{\mathrm{90}}(t-\tau_l)\right)
\end{equation}
In above, $x_{\mathrm{RF}}(t)$ denotes the PA output signal, $\tau_l$ is the RF delay of the $l^{th}$ tap, $\alpha_l$ and $\theta_l$ refer to the adjustable amplitude and phase values of the $l^{th}$ vector modulator, and
$x_{\mathrm{RF}}^\mathrm{0}(t)$ and $x_{\mathrm{RF}}^\mathrm{90}(t)$ denote the $0^{\circ}$ and $90^{\circ}$ phase-shifted tap signals inside a vector modulator, respectively. 
In the RF canceller design, we set the maximum delay $\tau_3$ to 10~ns which corresponds to round 3~m equivalent distance. This way, we can still facilitate modeling frequency selectivity in the direct SI coupling channel, while essentially avoid cancelling any echoes from true targets that are located more than 1.5~m away.

\begin{figure}[t]
    \centering
    \begin{subfigure}[b]{1\columnwidth}
        \centering
        \includegraphics[width=1.0\linewidth]{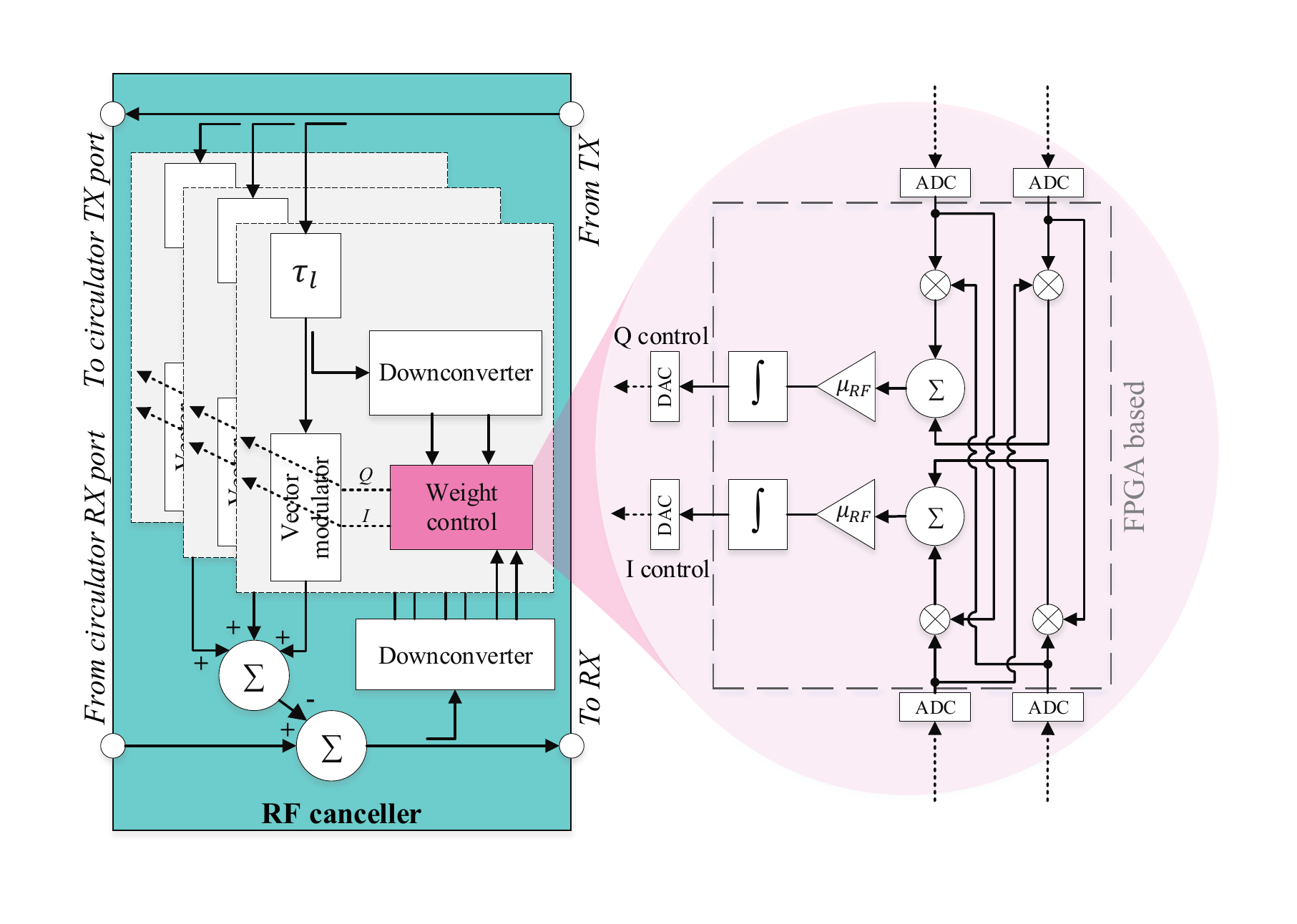}
        \caption{$ $}
    \end{subfigure}
    \hfill
    \vspace{1mm}
    \begin{subfigure}[b]{1\columnwidth}
        \centering
        \includegraphics[width=0.7\linewidth]{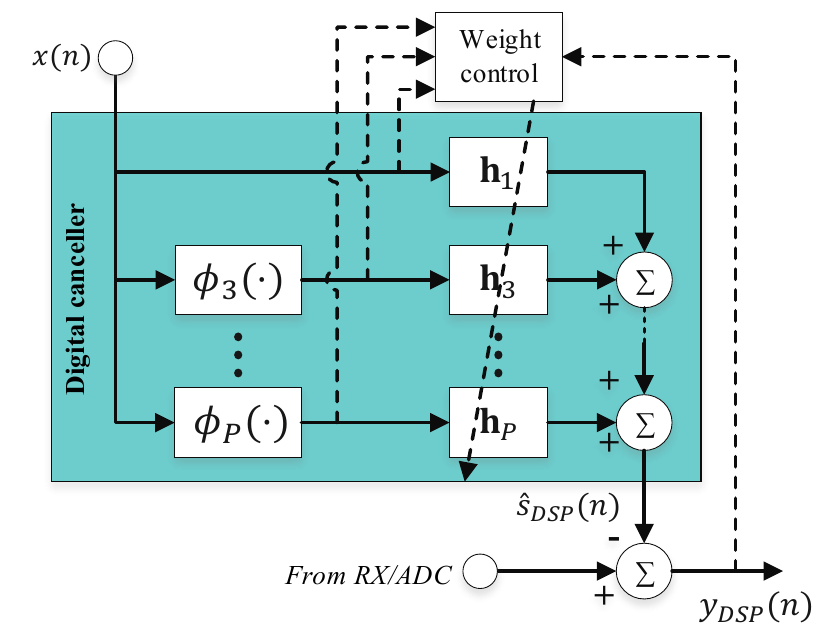}
        \caption{$ $}
    \end{subfigure}
    \caption{\quad In (a), the RF canceller block-diagram and self-adaptive weight control are shown. In (b), a general illustration of the nonlinear digital canceller is given.
    }
    \label{fig:diagram_cancellers}
\end{figure}

To accurately reduce the direct SI component power at the RF canceller output, proper real-time control of the amplitude and phases value, $\alpha_l$ and $\theta_l$, $l=1,2,3$, is essential. We have implemented a digital control system where complex I/Q downconverted observations of the tapped PA outputs and the RF canceller output, denoted by $x_{\mathrm{RF}}^{\mathrm{IQ}}(t)$ and $y^{\mathrm{IQ}}_{\mathrm{RF}}(t)$, respectively, are utilized. Building on gradient-based learning \cite{7146163}, implemented digitally on an FPGA, and combining the amplitudes and phases into corresponding complex coefficients $c_l=\alpha_l e^{j\theta_l}$ inside the digital control system, the RF canceller parameter learning can be expressed as
\begin{equation}
\label{RF_canceller_2}
    c_l \leftarrow c_l + \mu_{\mathrm{RF}} \int \left[x_{\mathrm{RF}}^{\mathrm{IQ}} (t-\tau_l) \right]^*  y^{\mathrm{IQ}}_{\mathrm{RF}}(t)dt
\end{equation}
where $\mu_{\mathrm{RF}}$ is the learning step size, $[\cdot]^*$ refers to complex conjugation, and the integration is performed digitally through the corresponding sampled I/Q signals. 
The digital control and tracking of the RF canceller weights is shown by measurements, in Section IV, to provide more than 50~dB of active RF cancellation gain with an example instantaneous bandwidth of 40~MHz. A block-diagram illustrating the overall RF canceller entity and its digital control system is shown in Fig.~\ref{fig:diagram_cancellers}(a).

Additionally, the residual direct SI  after the RF canceller can and must be further suppressed towards the receiver noise floor by using a digital cancellation stage. Inspired by our earlier work in \cite{7756408}, \cite{6810482}, we adopt a nonlinear digital canceller such that potential TX/RX RF nonlinearities can also be suppressed. Assuming a memory polynomial-based nonlinear processing approach, the corresponding digital cancellation signal can be expressed as 
\begin{equation}
\label{Digital_canceller_1}
    \hat{s}_{\mathrm{DSP}}(n) = \sum_{ \substack{p = 1 \\ p~\text{odd}}}^{P}  \sum_{k = -M_{1} }^{M_{2}} h_{p,k} \phi_p (x(n-k)) 
\end{equation}
where $x(n)$ denotes the digital I/Q transmit samples, $\phi_p (x(n)) = \left |  x(n) \right |^{p-1}x(n)$, $p=1,3,\dots,P$, are the nonlinear basis functions up to order $P$, and $h_{p,k}$ denote the adjustable filter parameters with $M_1$ and $M_2$ precursor and postcursor memory taps, respectively, per each nonlinearity order $p$. For parameter estimation, block least-squares (LS) or adaptive filtering methods such as least-mean-square (LMS) or recursive least-squares (RLS) can in general be adopted \cite{7565192}, \cite{7756408}, seeking to minimize the cancelled signal power.

As acknowledged already in prior literature \cite{7565192}, \cite{7756408}, the different nonlinear basis functions $\phi_p (x(n))$, $p=1,3,\dots,P$, are largely correlated which, in turn, complicates parameter estimation, particularly when gradient based methods are adopted. Instead of explicitly orthogonalizing the basis functions, as done in prior work \cite{7565192}, \cite{7756408}, we adopt a novel self-orthogonalizing learning rule in this work, expressed as
\begin{equation}
\label{DSP_canceller_learning}
    \mathbf{h} \leftarrow \mathbf{h} + \mu_{\mathrm{DSP}}\mathbf{C}^{-1}\mathbf{u}(n)y_{\mathrm{DSP}}(n)
\end{equation}
where $\mathbf{h}$ is a column-vector collecting the canceller coefficients $h_{p,k}$, $\forall p,k$, $\mu_{\mathrm{DSP}}$ refers to the learning step-size, and $\mathbf{C}$ denotes the correlation matrix of the basis function samples which can be precomputed. Additionally, $\mathbf{u}(n)$ refers to the basis function samples used to calculate the cancellation signal sample $\hat{s}_{\mathrm{DSP}}(n)$ as shown in (\ref{Digital_canceller_1}) while $y_{\mathrm{DSP}}(n)$ denotes the corresponding sample at the digital canceller output. This approach greatly reduces the computational complexity of the main digital cancellation path, as the explicit basis function orthogonalization utilized in \cite{7565192}, \cite{7756408} is avoided. A principal illustration of the digital cancellation system is shown in Fig.~\ref{fig:diagram_cancellers}(b).

Similar to the RF canceller, the memory length of the digital cancellation processing must be chosen with care, in order to avoid suppressing the reflections from true targets.
Particularly, the amount of the postcursor taps $M_2$ used in the digital canceller is directly related to the detectable radar range. In our implementation, the digital canceller runs at a sample rate of 240 MHz and contains $M_1=5$ precursor and $M_2=5$ postcursor taps. These correspond to minimum detectable distance of ca. 3~m for true targets which is well inline with the RF canceller parameterization.

\begin{figure}[t]
    \centering
    \begin{subfigure}[b]{1\columnwidth}
        \centering
        \includegraphics[width=\linewidth]{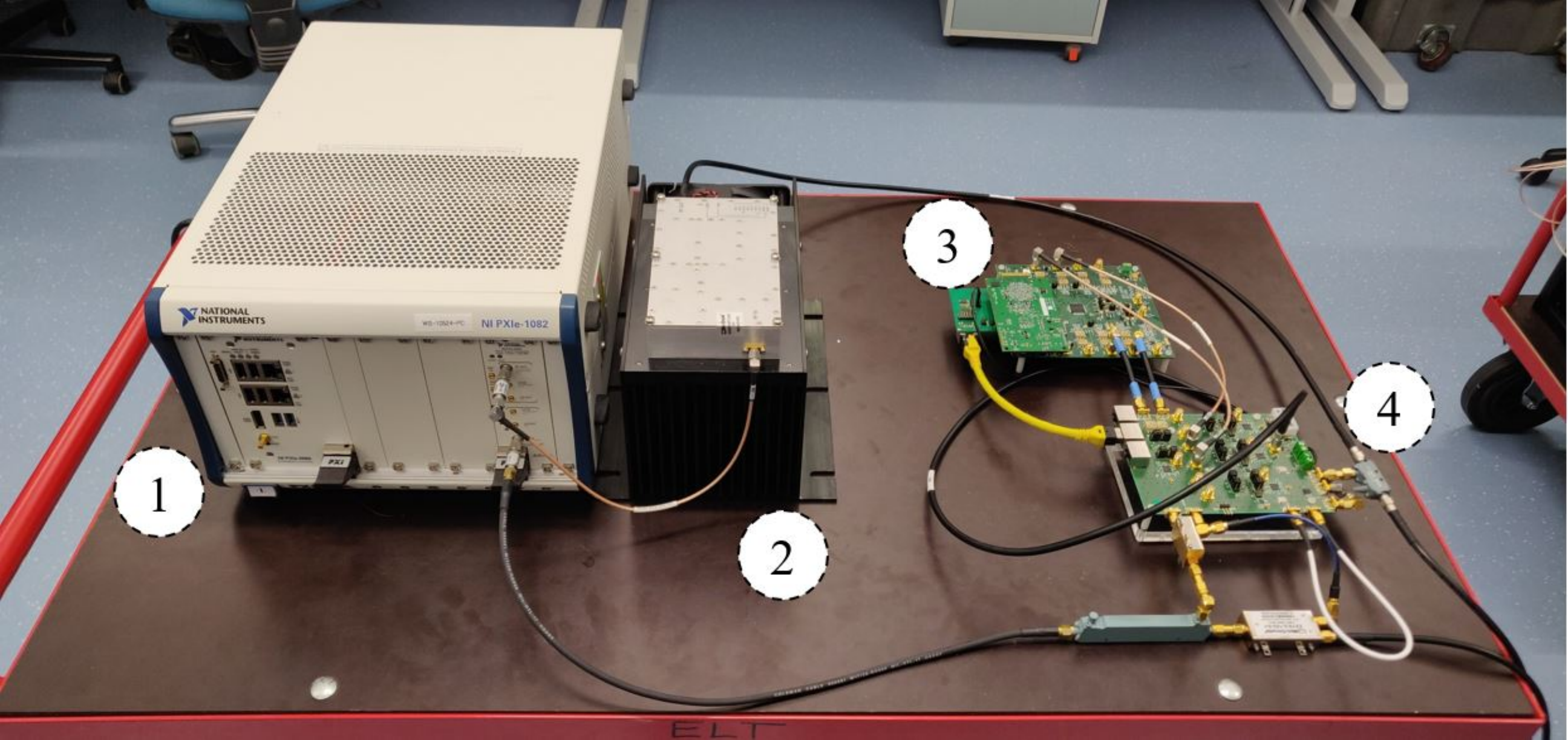}
        \caption{$ $}
        \label{fig:setup_equipment}
    \end{subfigure}
    \hfill
    \vspace{1mm}
    \begin{subfigure}[b]{1\columnwidth}
        \centering
        \includegraphics[width=\linewidth]{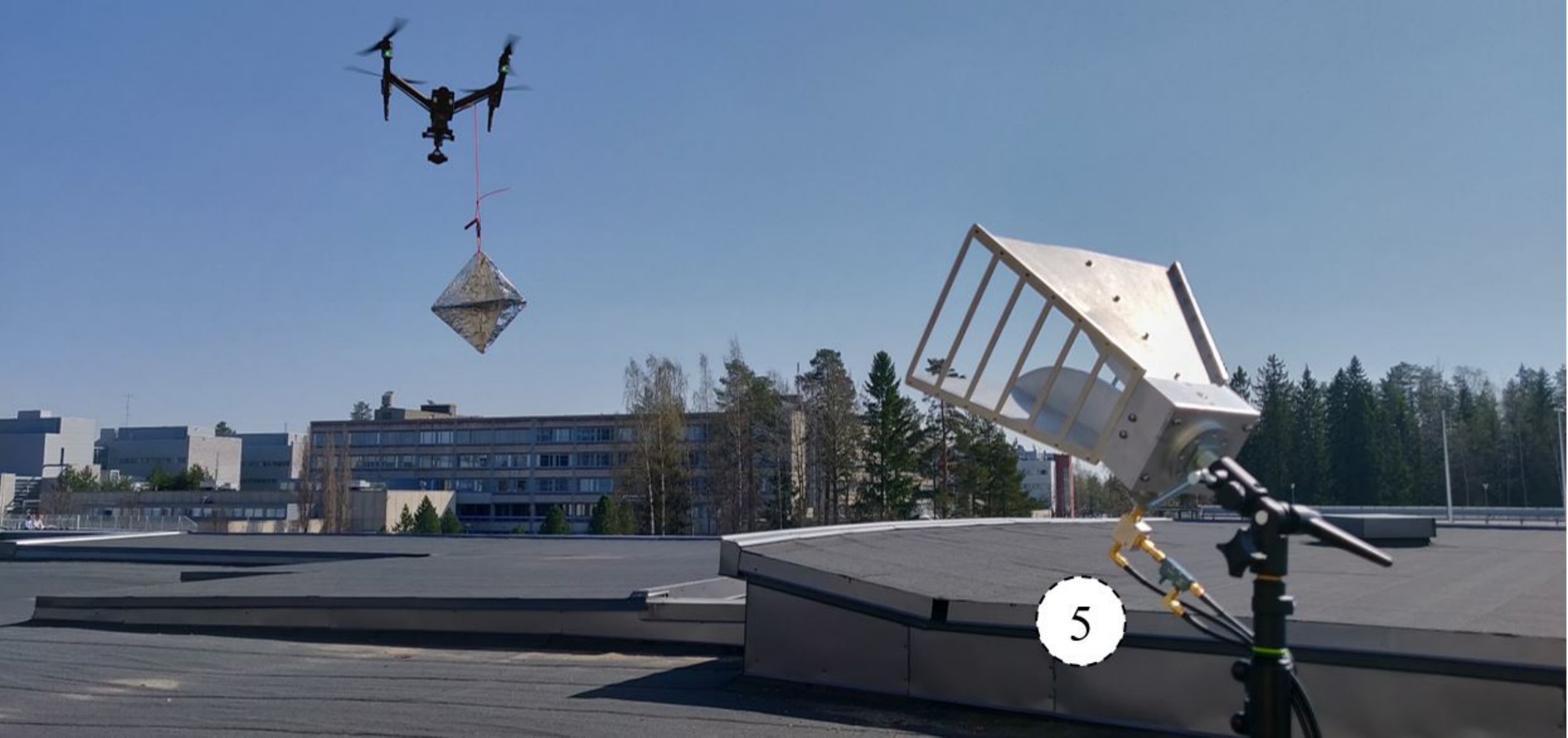}
        \caption{$ $}
        \label{fig:setup_antennas}
    \end{subfigure}
    \caption{\quad In (a), the main equipment used in the RF measurements are shown, namely {\normalsize \textcircled{\footnotesize{1}}}
  vector signal transceiver (VST), {\normalsize \textcircled{\footnotesize{2}}}  external power amplifier, {\normalsize \textcircled{\footnotesize{3}}} the developed RF canceller and {\normalsize \textcircled{\footnotesize{4}}} its digital control board. In (b), the directive TX/RX horn antenna and circulator {\normalsize \textcircled{\footnotesize{5}}} are shown, together with a drone (DJI Insipire-2) carrying a retro-reflector.}
    \label{fig:setup}
\end{figure}

\begin{figure}[t]
    \centering
    \begin{subfigure}[b]{1\columnwidth}
        \centering
        \includegraphics[width=0.95\columnwidth]{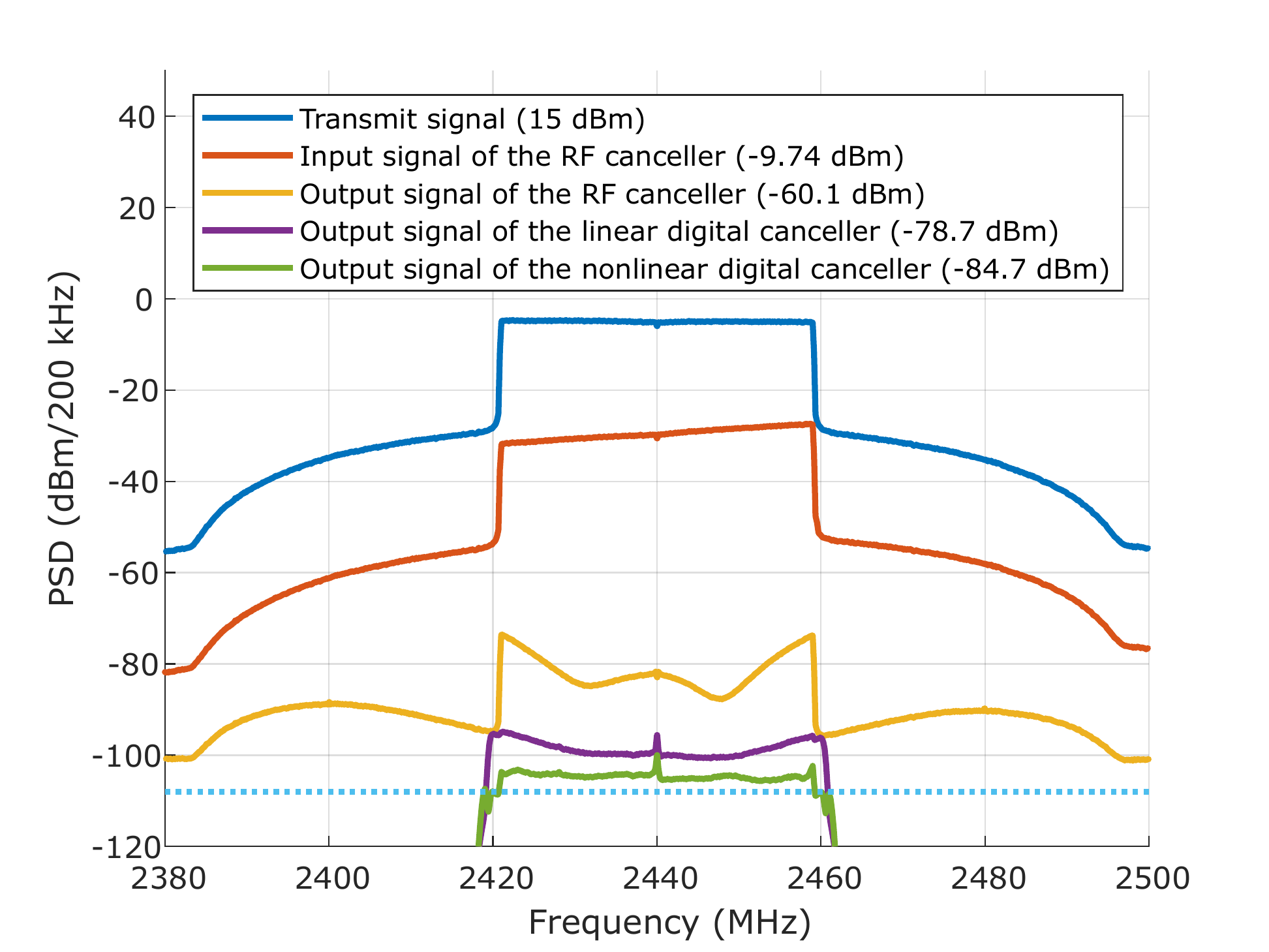}
        \caption{$ $}
        \label{fig:canceller_spectrum}
    \end{subfigure}
    \hfill
    \vspace{1mm}
    \begin{subfigure}[b]{1\columnwidth}
        \centering
        \includegraphics[width=0.95\columnwidth]{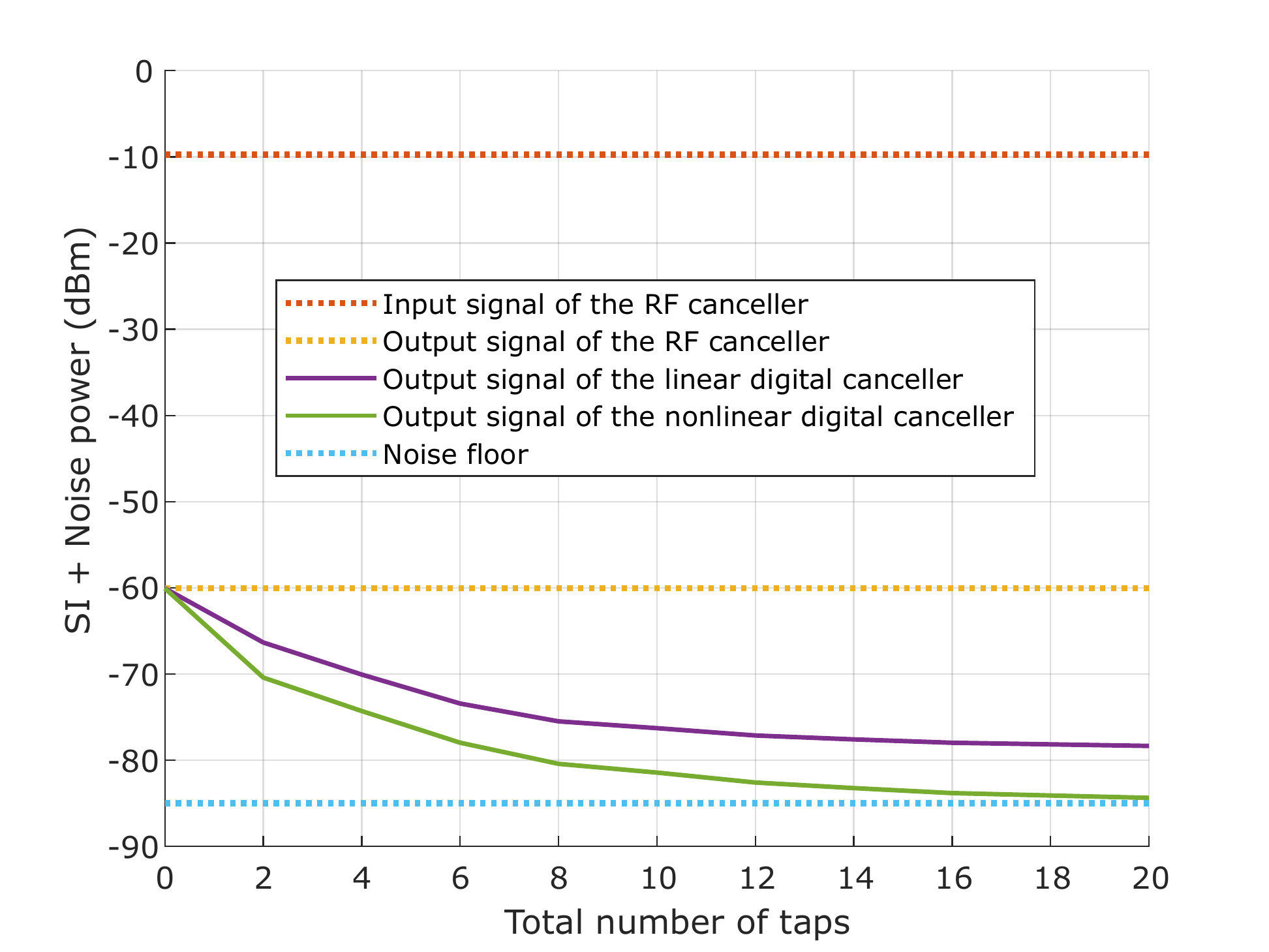}
        \caption{$ $}
        \label{fig:canceller_taps}
    \end{subfigure}
    \caption{\quad Overall measured TX-RX isolation performance with 40 MHz NR waveform at 2.44 GHz for (a) 5 precursor and 5 postcursor taps and (b) varying total number of taps in digital cancellation. The order of the nonlinear digital canceller is $P=11$, while the linear digital canceller corresponds to $P=1$.
    }
    \vspace{2mm}
    \label{fig:canceller_isolation}
\end{figure}

\begin{figure*}[t!]
    \centering
    \begin{subfigure}[t]{0.32\textwidth}
        \centering
        \includegraphics[width=1\columnwidth]{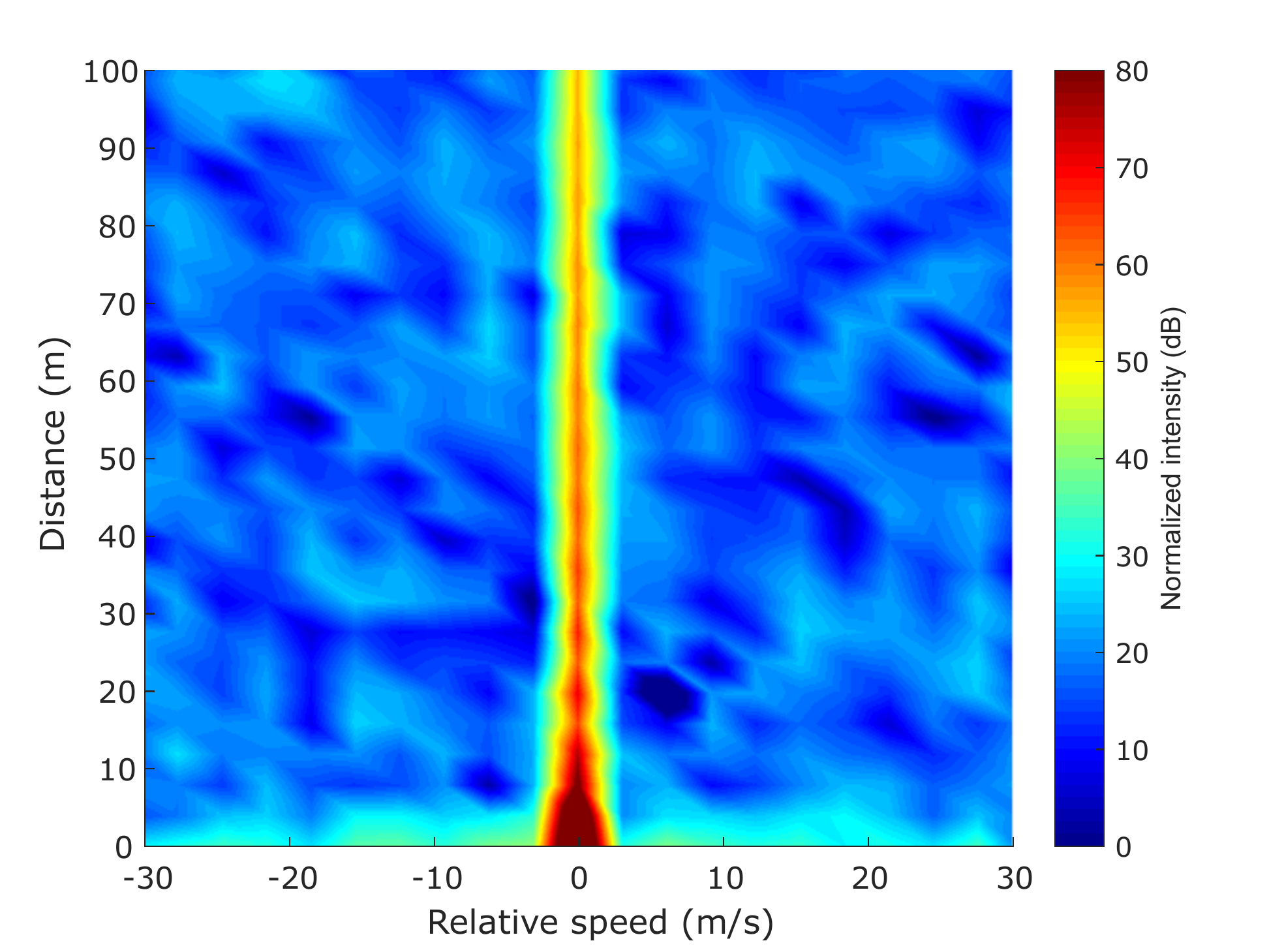}
        \caption{}
        \label{fig:meas_period_drone_ref}
    \end{subfigure}%
    ~ 
    \begin{subfigure}[t]{0.32\textwidth}
        \centering
        \includegraphics[width=1\columnwidth]{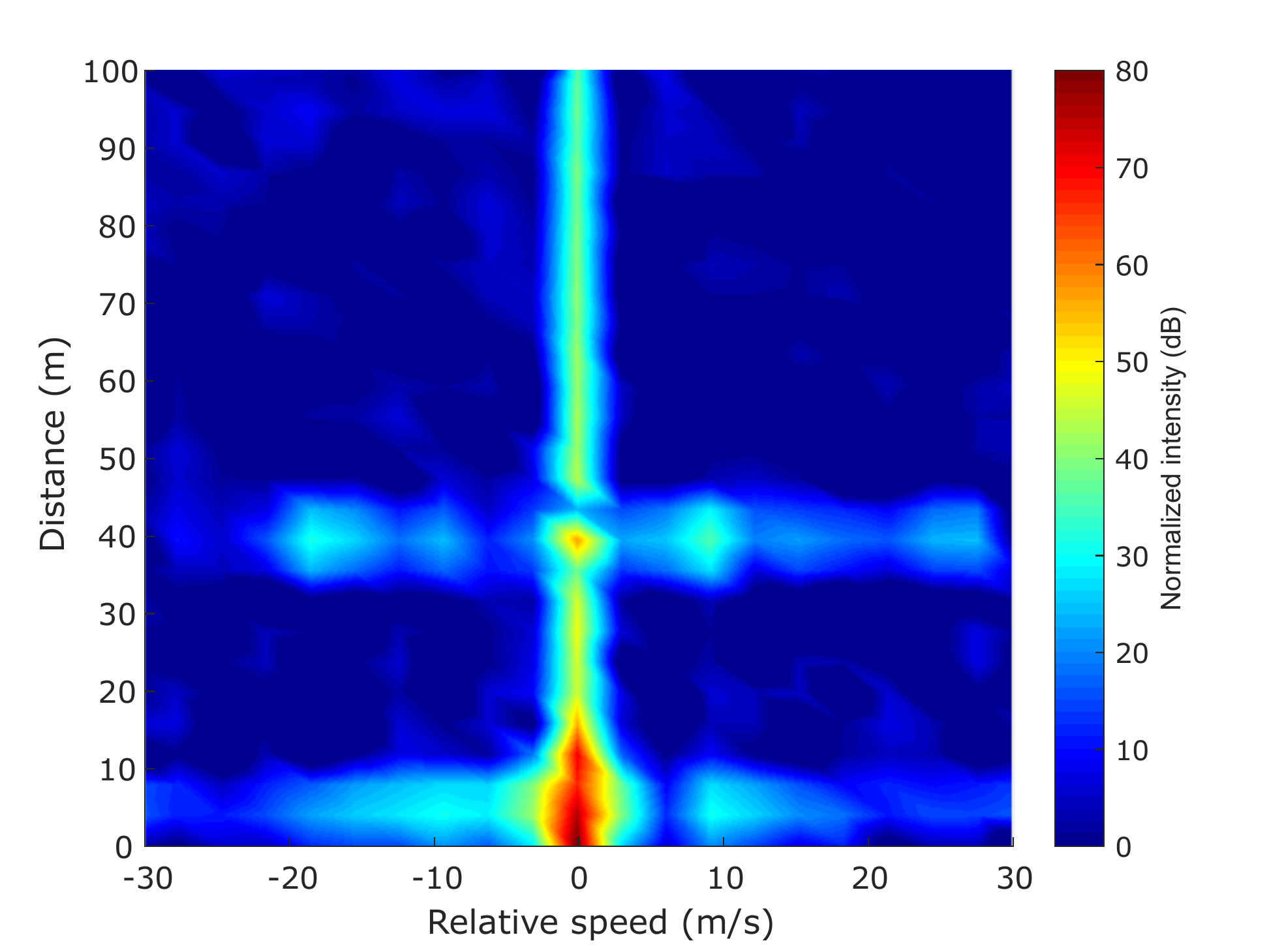}
        \caption{}
        \label{fig:meas_period_drone_RF}
    \end{subfigure}
    ~ 
    \begin{subfigure}[t]{0.32\textwidth}
        \centering
        \includegraphics[width=1\columnwidth]{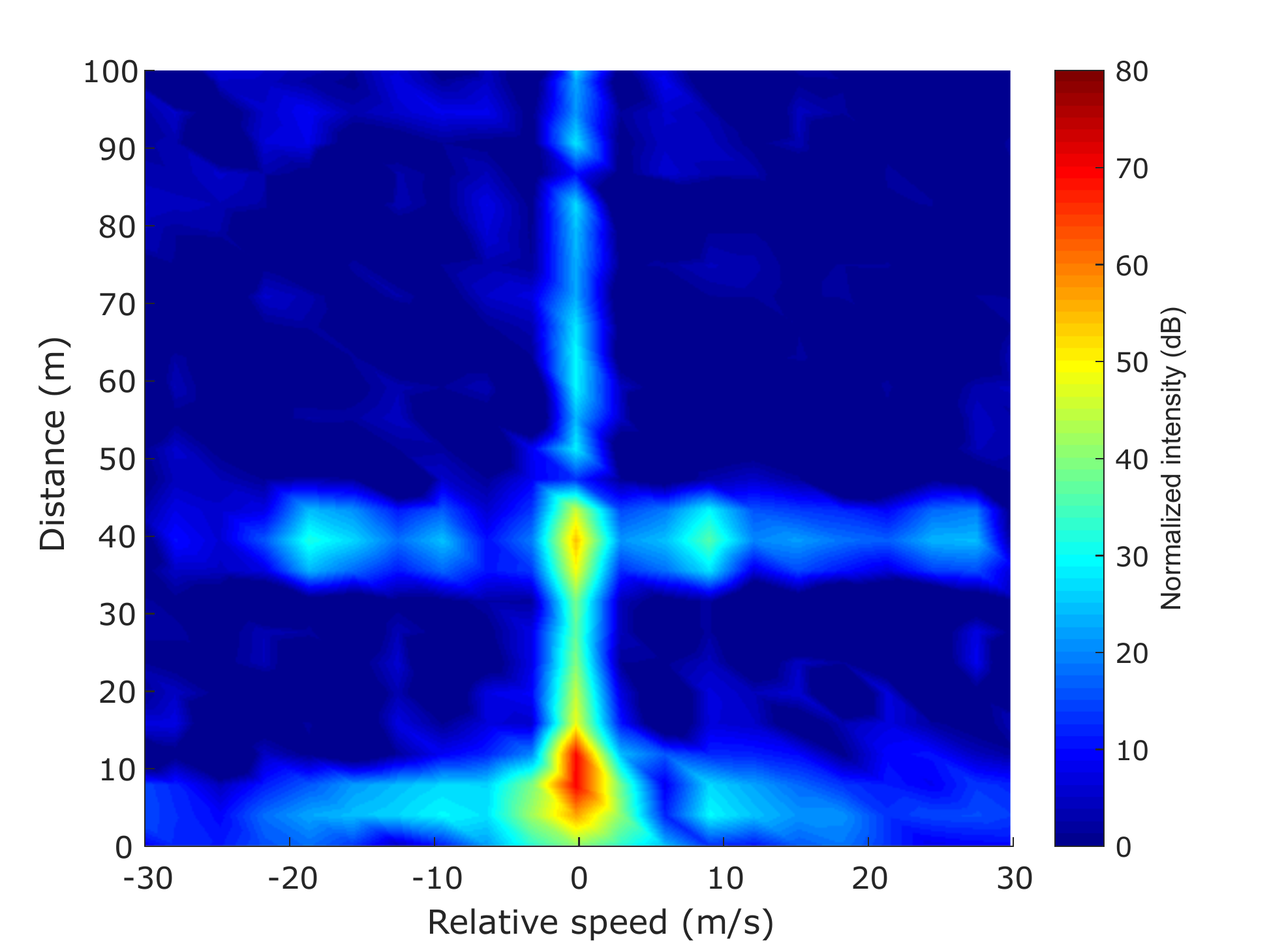}
        \caption{}
        \label{fig:meas_period_drone_Digital}
    \end{subfigure}
     \centering
    \begin{subfigure}[t]{0.45\textwidth}
        \centering
        \includegraphics[width=1\columnwidth]{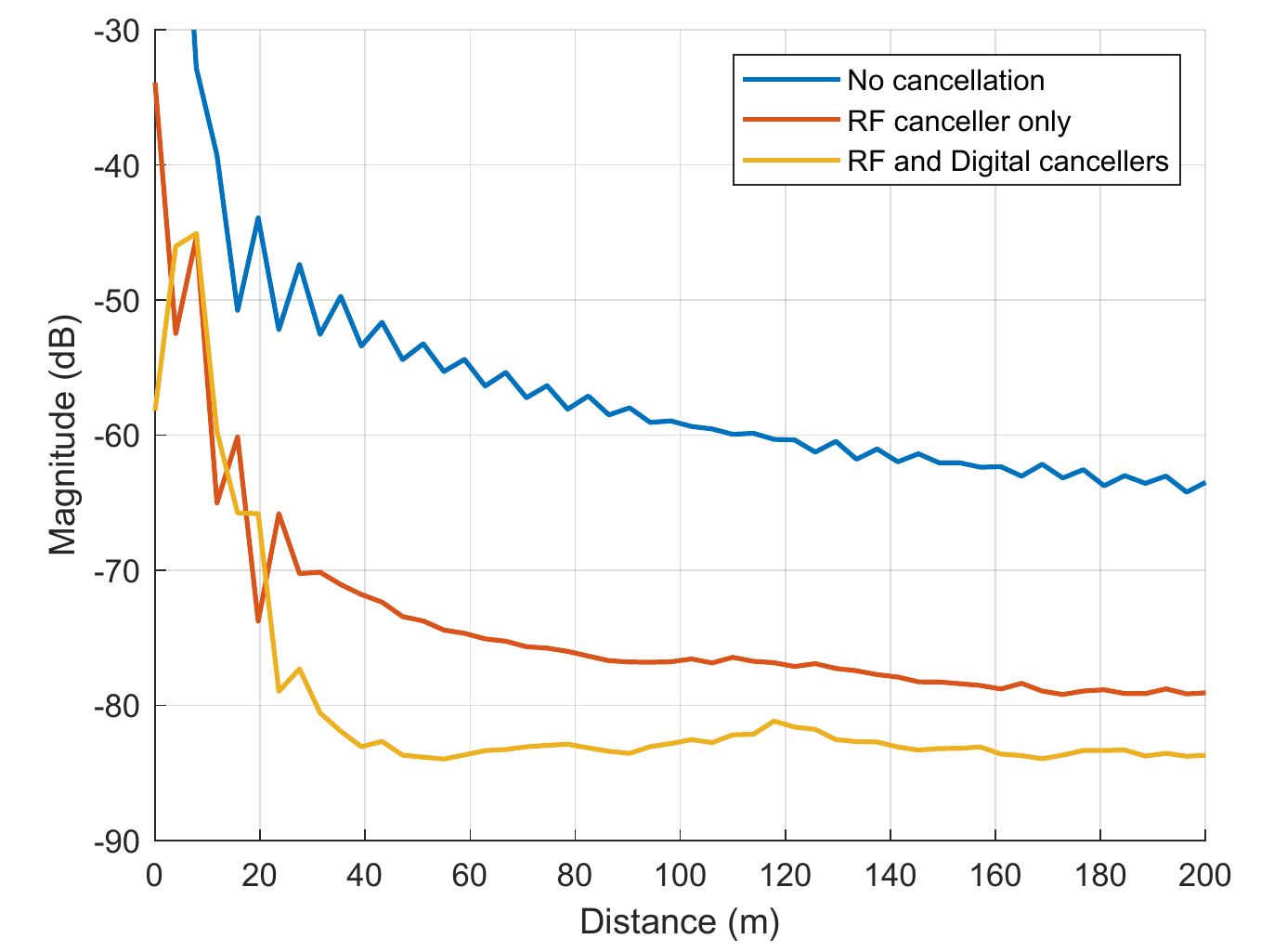}
        \caption{}
        \label{fig:SI_ref}
    \end{subfigure}%
    ~ 
    \begin{subfigure}[t]{0.45\textwidth}
        \centering
        \includegraphics[width=1\columnwidth]{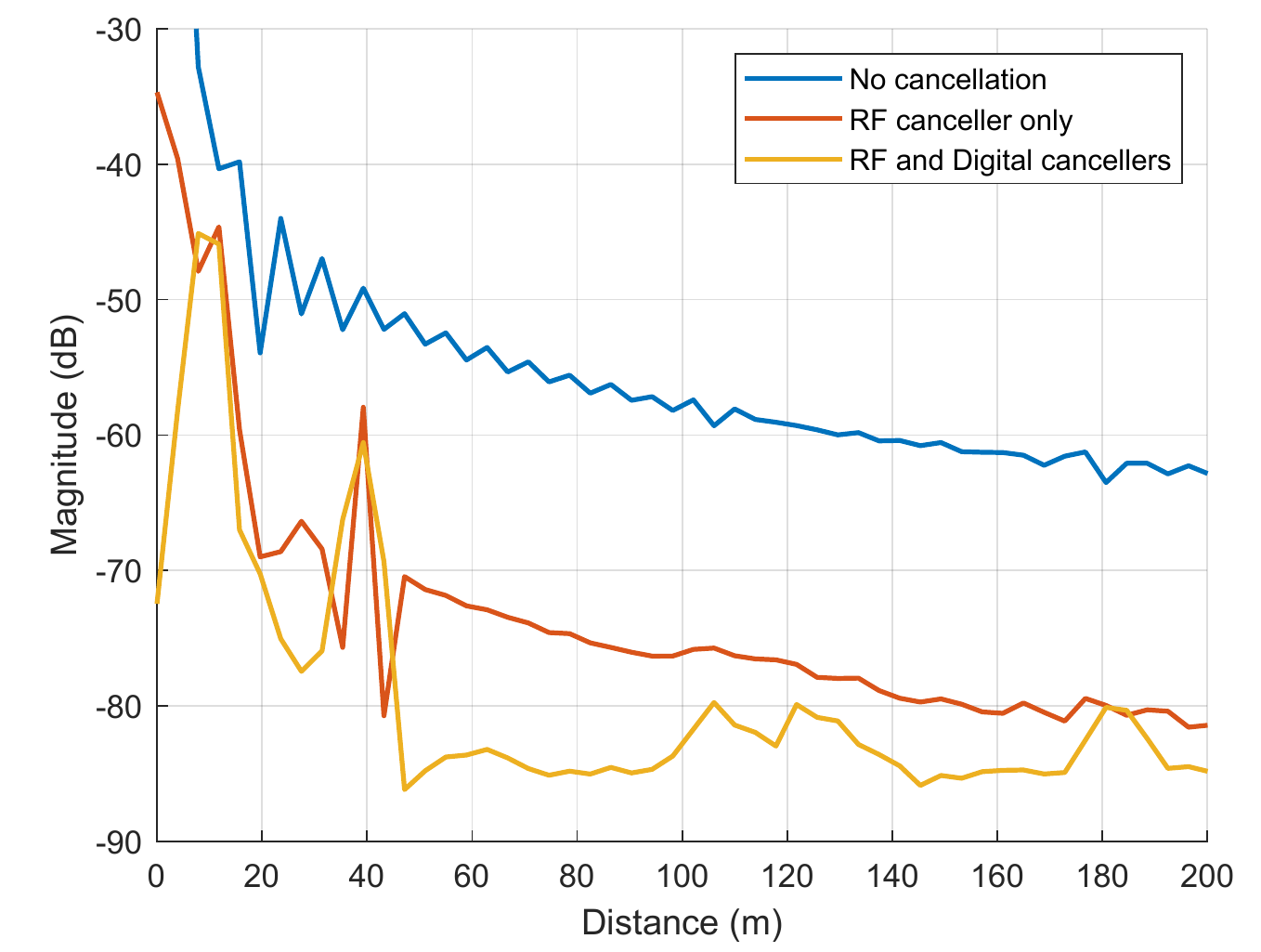}
        \caption{}
        \label{fig:SI_with_drone}
    \end{subfigure}
    \caption{\quad Measured radar images with a static airborne drone at a distance of 40~m, (a) without any cancellation, (b) with RF cancellation only, and (c) with both RF and digital cancellation. In (d) and (e), distance estimation profiles of reference measurement (no target) and drone measurement are shown, respectively, illustrating also the additional leakage suppression gain from the RF and digital cancellers.}
    \label{fig:measurement_drone}
\end{figure*}


\section{Implementation and Measurement Results}
\label{sec:implementation}

\subsection{Measurement Setup}
The basic elements of the measurement setup are illustrated in Fig.~\ref{fig:setup}, building on the National Instruments PXIe-1082 vector signal transceiver (VST) implementing the basic RF transmitter and receiver functionalities. An external power amplifier (PA) with a gain of 42~dB is also utilized, and all the measurements are carried out at the 2.4~GHz ISM band with ca. $+20$~dBm transmit power. The RF canceller and its digital control functionalities are implemented on separate boards, properly interfaced to the VST and the external PA. A directive LB-880 horn antenna with 10 dBi of antenna gain is adopted and shared between the TX and RX through a circulator (JQL~JCC2300T2500S6), which together provide some 25~dB of passive TX--RX isolation, partially due to antenna reflection and partially due to finite isolation between the circulator TX and RX ports.
The digital SI cancellation and radar processing are implemented in the VST host processing environment.

In all the measurements, a 5G NR waveform with a subcarrier spacing of 30~kHz and 40~MHz channel bandwidth is adopted (cf.\ Table I), with FFT/IFFT size of 2048 implying a core I/Q time-domain sample rate of $2048 \times 30~\mathrm{kHz}=61.44~\mathrm{MHz}$. Further PAPR reduction and OFDM symbol windowing are also implemented in the transmitter chain. The digital front-end sample rate is 240~MHz at which the digital SI canceller is also running. The order of the digital canceller is either $P=1$ (linear digital canceller) or $P=11$ (nonlinear digital canceller). After the digital cancellation, the digital received signal is down-sampled by $125/32$ to the fundamental I/Q sample rate of $61.44~\mathrm{MHz}$. Additionally, the Hamming window \cite{richards2010principles} is used in the periodogram calculations in all the measurements.

\subsection{Measured TX--RX Isolation}
Before going to the actual radar experiments, the achievable total TX--RX isolation of the developed demonstrator system incorporating the RF and digital cancellation solutions is measured and illustrated. An example of the power spectral densities in different stages of the transceiver system is shown in Fig. \ref{fig:canceller_isolation}(a), measured in an anechoic chamber with antenna connected, while no actual targets are present. We can observe that the RF canceller provides more than 50~dB of the direct SI suppression, while being then further complemented by the digital canceller such that an overall isolation of ca. 100~dB is reached. We also note that the total isolation provided by the circulator and the active RF canceller is some 75~dB, which is very essential to prevent receiver saturation, especially with larger transmit powers. Such 75~dB RF isolation is already close to the isolation provided by high-quality base-station duplex filters in frequency-division duplexing (FDD) based networks \cite{2018Dahlman5G}, hence potentially facilitating simultaneous transmission and reception for transmit powers up to +40~dBm, or even +46~dBm, from the receiver linearity and saturation point of view.

The figure also clearly shows that the linear digital canceller is limited by the RF nonlinearities, while the nonlinear digital canceller can efficiently suppress also the nonlinear leakage. This is one complementary ingredient that the radar domain based direct leakage cancellers cannot most likely facilitate. To better assess the impact of the digital canceller parameterization, in terms of the number of pre-cursor and post-cursor taps, Fig.~\ref{fig:canceller_isolation}(b) shows the integrated passband powers for varying numbers of total taps in the digital canceller, while also comparing the linear and nonlinear digital cancellers. We can conclude that increasing the amount of the canceller taps, and hence complexity, beyond the case of 5+5 taps provides only a marginal performance improvement. Hence, in all the following experiments, the 5+5 configuration is adopted.

\subsection{Sensing Static Targets under SI}

Next, we continue with the RF measurements through actual outdoor experiments containing real targets. As a first example, we utilize a commercial drone flying at around 38~m height above the OFDM radar system holding a multi-faceted reflector as illustrated already in Fig.~\ref{fig:setup}(b).
The drone is deliberately staying at fixed coordinates, thus representing a static airborne target at an overall distance of some 40~m.
Fig.~\ref{fig:measurement_drone}(a)
presents the corresponding measured radar image, yet without any RF or digital cancellation, meaning that only the passive circulator isolation is utilized. It can clearly be observed, similar to the synthetic data based radar images in Fig.~\ref{fig:SI_analysis}, that the SI corresponds to a very strong static target located essentially at zero distance and zero velocity, and also that the sidelobes are strong up to large distances. 

Then, Figs.~\ref{fig:measurement_drone}(b) and (c) show how the OFDM radar-based range--Doppler maps are improving when the presented RF canceller, in (b), or both the RF and digital cancellers, in (c), are also applied. Opposed to Fig.~\ref{fig:measurement_drone}(a),  a clear static target is now seen, at around the distance corresponding to the drone. It is also noted that even though the drone is remaining at a fixed distance with zero velocity during the measurements, the radar image shows some additional peaks at the same distance of 40~m, but non-zero velocity. These are stemming from the propellers' rotation, i.e., reflect the micro-Doppler phenomenon. Overall, the measurement-based observations are well inline with the simulated results discussed along Fig.~\ref{fig:SI_analysis}. 

To further assess and visualize the impact of the cancellers, we carry out additional reference measurements such that the antennas are tilted towards the sky and no actual targets are present. This way, the receiver is essentially observing and processing only the SI and some residual reflections and clutter from the surroundings. The  range profile of such reference measurements is shown in Fig.~\ref{fig:measurement_drone}(d) while the corresponding range profile of the drone measurements is shown in Fig.~\ref{fig:measurement_drone}(e). By comparing the range profiles, we can clearly observe that there is a relatively strong reflection at a distance of ca. 10~m, caused most likely be the antenna sidelobe and close-by building surface. Additionally, the figures clearly illustrate the improved dynamic range offered by the RF and digital cancellers. When both the RF and digital cancellers are adopted, also reflections from high-rise buildings at distances of some 120~m and 180~m can be observed, reflecting the increased dynamic range.

\begin{figure}[t]
        \centering
        \includegraphics[width=0.98\columnwidth]{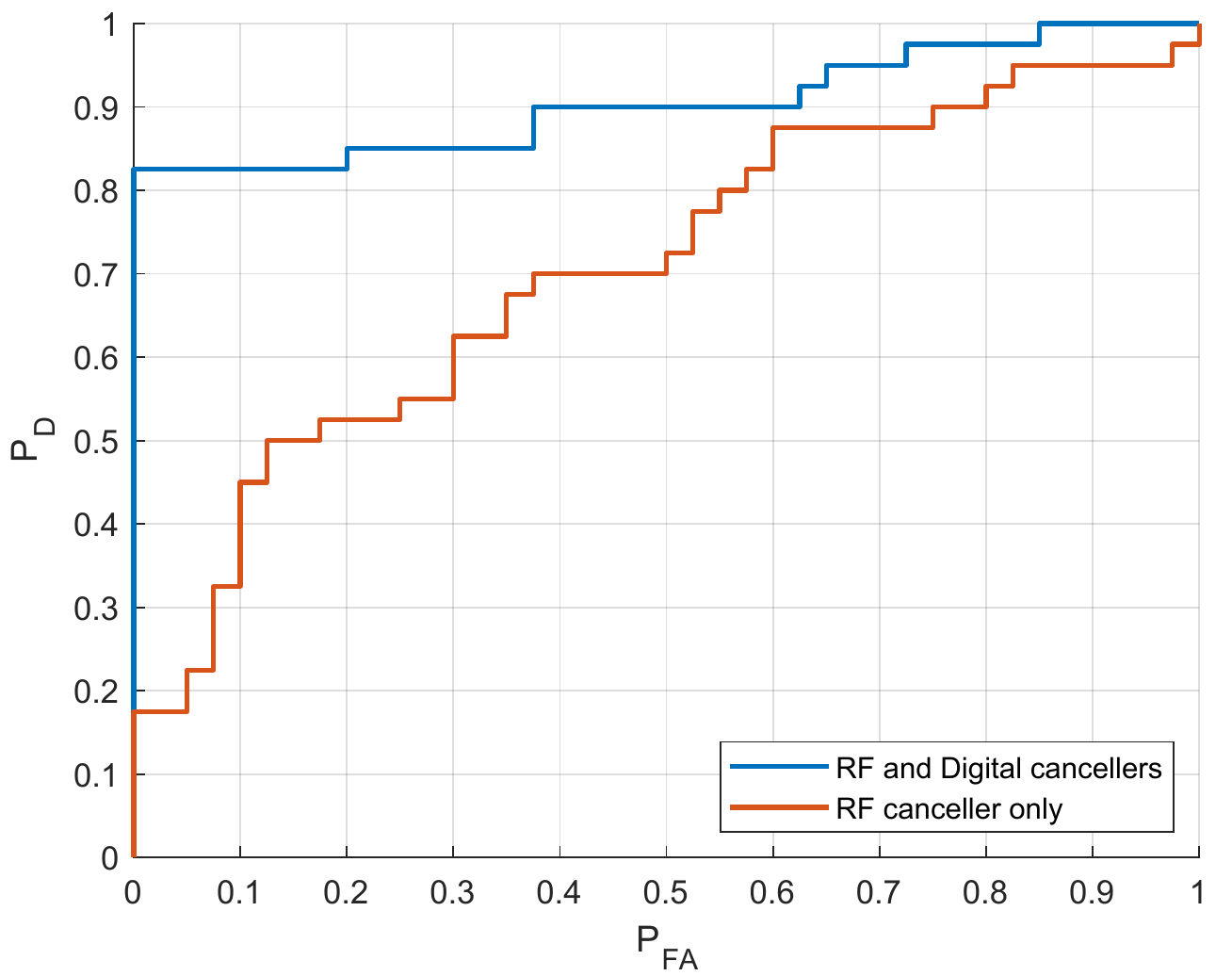}
        \caption{\quad Measured receiver operating characteristic (ROC) curves in the drone scenario, with RF canceller only and with both RF and digital cancellers. 
        }
        \label{fig:ROC_drone}
\end{figure}

Next, we assess the impact of the RF and digital cancellers on the actual drone detection, through the measured data. We first measure 100 radar images without the drone (hypothesis $H_0$) and another 100 radar images with the drone (hypothesis $H_1$). Then, we calculate the empirical distributions of the radar images for both data sets, covering $3 \times 3$ pixels of the periodograms at and around the drone location. We then vary the detection threshold $T_\mathrm{th}$ and evaluate the probability of detection and probability of false alarm through the measured empirical distributions. These compose the measured receiver operating characteristics (ROC) curve, shown visually in Fig.~\ref{fig:ROC_drone}. The improvement provided by the digital canceller, compared to RF canceller only, can be clearly observed.

\subsection{Sensing Moving Targets under SI}
\begin{figure}[t]
    \centering
    \begin{subfigure}[h]{1\columnwidth}
        \centering
        \includegraphics[width=1\columnwidth]{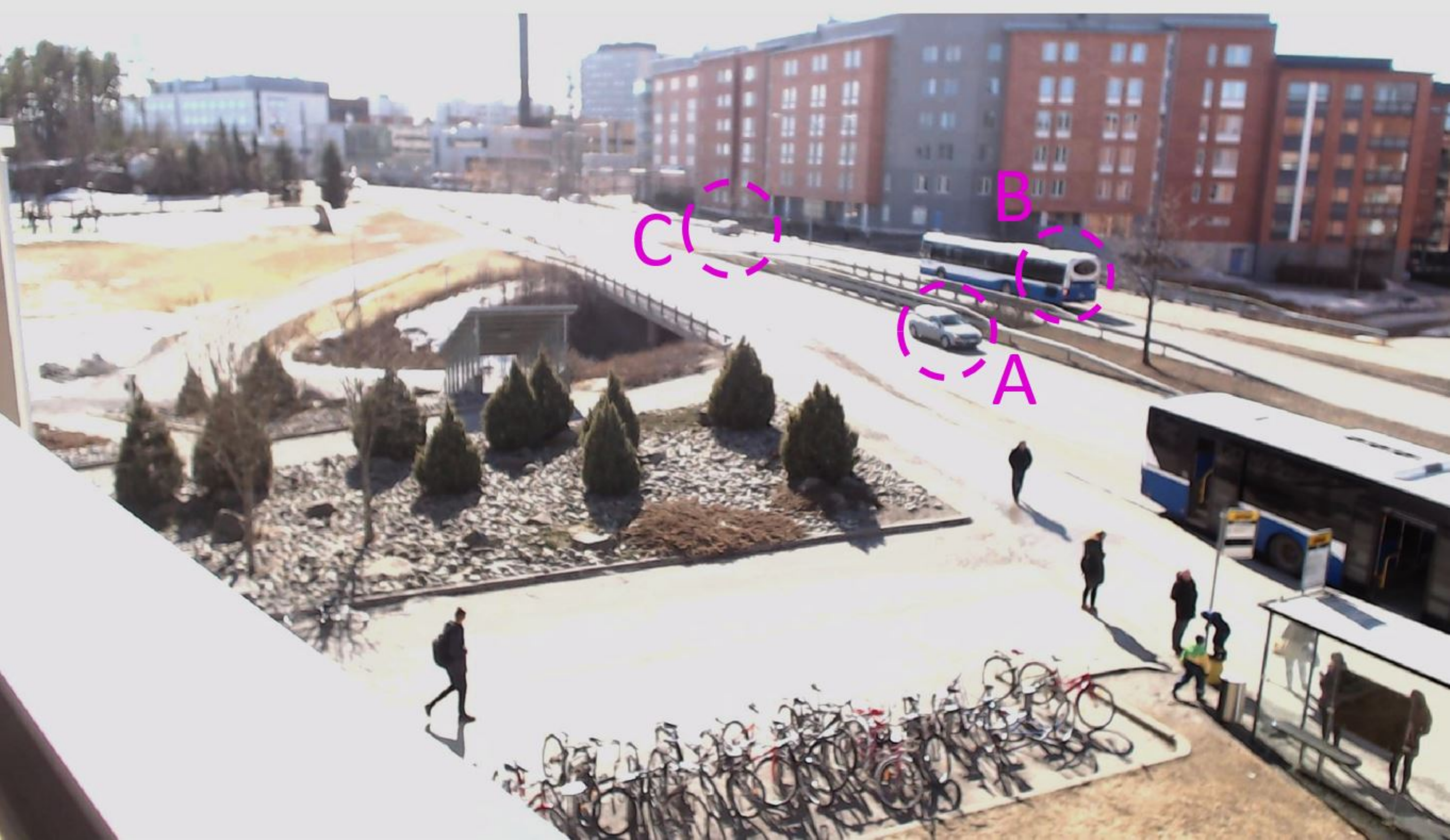}
        \caption{$ $}
        \label{fig:Meas_moving_image}
    \end{subfigure}
    \hfill
    \begin{subfigure}[h]{1\columnwidth}
        \centering
        \includegraphics[width=1\columnwidth]{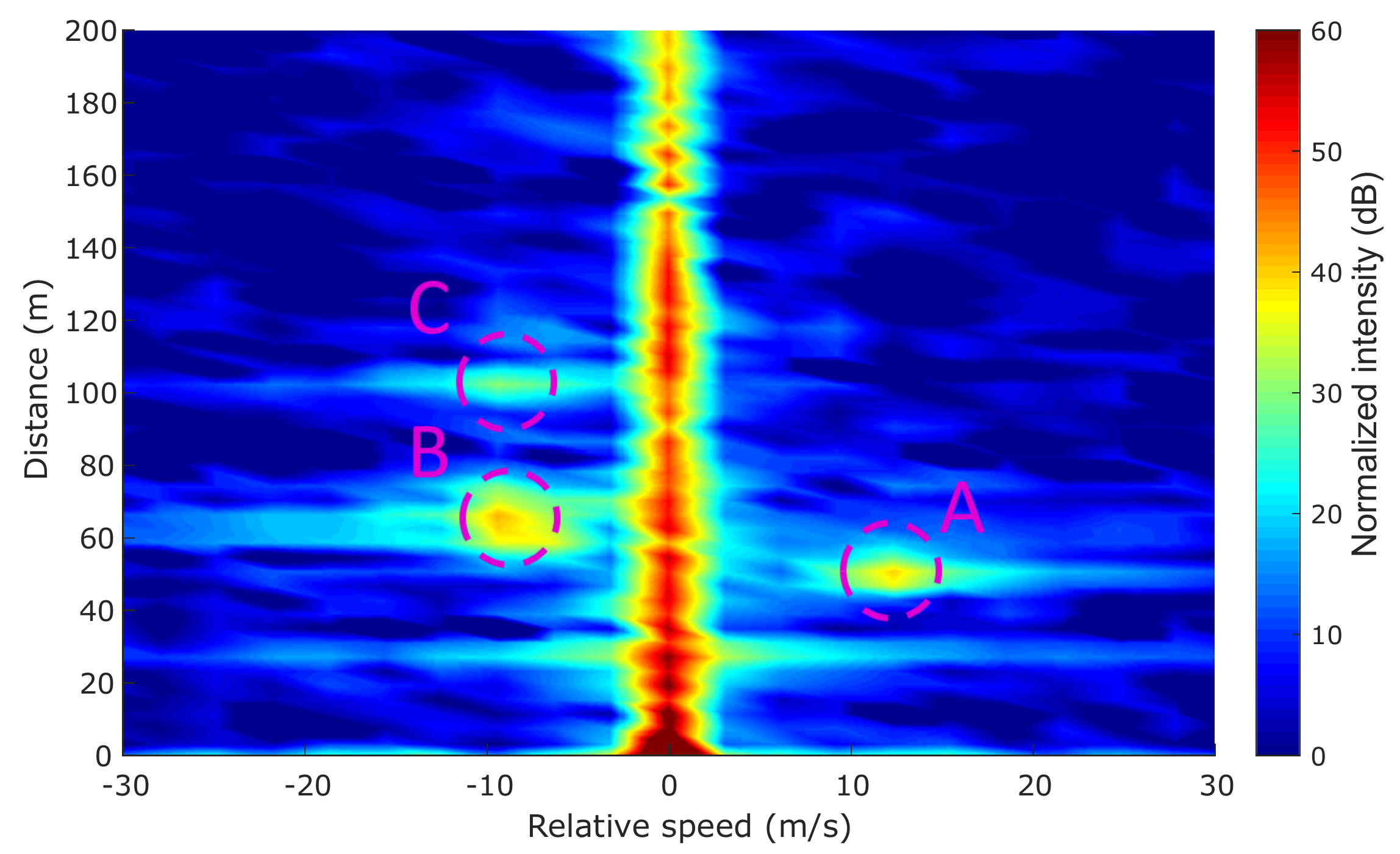}
        \caption{$ $}
        \label{fig:meas_moving_period_RF}
    \end{subfigure}
    
    \begin{subfigure}[h]{1\columnwidth}
        \centering
        \includegraphics[width=1\columnwidth]{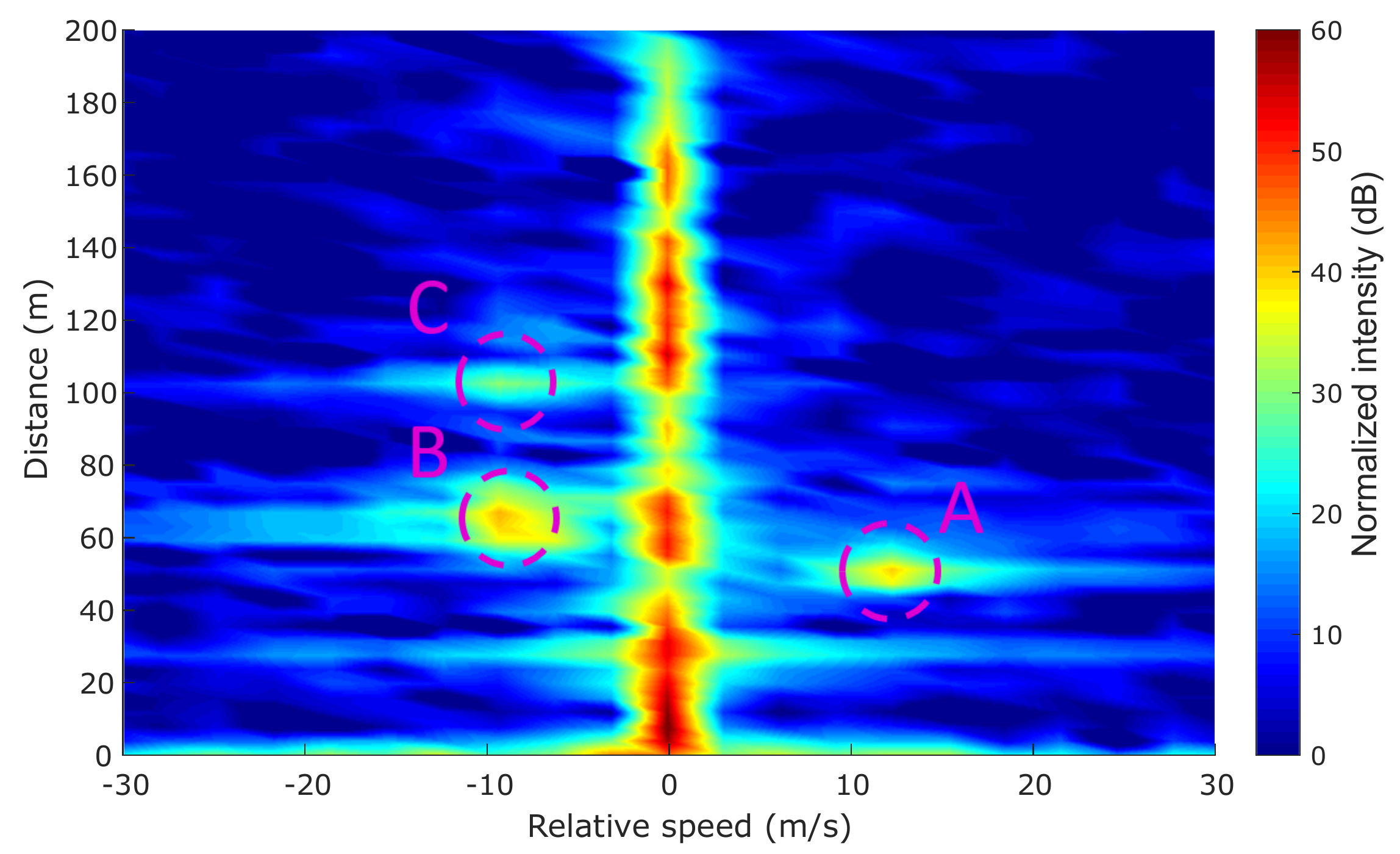}
        \caption{$ $}
        \label{fig:meas_moving_period_Digital}
    \end{subfigure}
    \caption{\quad (a) Vehicular measurement scenario with three moving cars $A$, $B$ and $C$ located at snap-shot distances of 51.1~m, 66.8~m and 102.1~m, respectively, driving along the road such that the velocity relative to the radar system is ca. 12~m/s, -9~m/s and -9~m/s, respectively. The corresponding measured radar images with (b) RF canceller only and (c) both RF and digital cancellers, using an NR waveform of 20~ms with channel bandwidth of 40~MHz and subcarrier spacing of 30~kHz are shown. Additional multimedia material available at  {\color{blue}\url{http://www.tut.fi/full-duplex/radar/NR40cars2.mp4}}
    }
    \label{fig:Meas_moving}
\end{figure}

The overall OFDM radar system concept comprising the canceller solutions is next demonstrated and evaluated in a typical vehicular traffic scenario with three moving cars.  
Fig.~\ref{fig:Meas_moving}(a)illustrates the measurement scenario where three vehicles labeled $A$, $B$ and $C$ are driving along a road towards the OFDM radar system at distances of some 50--110~m. The measurement equipment and the antenna are located in the third floor of one of the university buildings, located next to a road with 50~km/h speed limit, and the circulator based shared-antenna system is again adopted.

The corresponding measured radar images shown in Fig.~\ref{fig:Meas_moving}(b) with RF cancellation only, and in Fig.~\ref{fig:Meas_moving}(c) with both RF and digital cancellers, 
demonstrate that the proposed system is clearly able to sense all three moving vehicles despite the strong reflections coming from the surrounding buildings as well as the static clutter. The estimated snap-shot distances and velocities of the vehicles $A$, $B$ and $C$, building on one radar image calculated over two NR radio frames, i.e., 20~ms, are 51.1~m, 66.8~m and 102.1~m, with relative speeds of 12~m/s, -9~m/s and -9~m/s, respectively. As the radar transmit--receive system and the corresponding antenna unit are not placed in the main road direction, the estimated velocities correspond to the velocity projection along a line between the considered vehicle and the radar system. Estimating that the angle between the main vehicle trajectory and the radar system location is ca. 20$^\circ$, the absolute velocities of the three vehicles are between 30~km/h to 40~km/h, which are consistent with the speed limit of the road.

Additional multimedia content is attached along the submission and is also available at {\color{blue}\url{http://www.tut.fi/full-duplex/radar/NR40cars2.mp4}}. In this multimedia illustration, a video recording of the moving cars is shown, together with displaying a sequence of the corresponding radar images, side by side. Each radar image is calculated over one 20~ms NR radio frame, while the overall experiment corresponds to ca. 10 seconds of real time. The sequence of the radar images is illustrating that moving vehicles can be efficiently sensed, despite the static reflections and ground clutter.


\section{Conclusion}
\label{sec:conclusion}

In this paper, the use of the LTE and 5G NR downlink waveforms for radar/sensing purposes was addressed and studied, together with the fundamental hardware challenges related to transmitter--receiver (TX--RX) isolation and associated self-interference problem. First, frequency-domain radar processing building on the LTE or NR time--frequency resource grid, complemented with interpolation to account for missing samples due to the null subcarriers within the transmit waveform passband, was described. Assuming ideal TX--RX isolation, for reference, the achievable target detection as well as the corresponding range and velocity estimation performances of the considered method were then evaluated numerically. Particularly the large channel bandwidths supported by 5G NR were shown to provide good sensing performance already at below 6 GHz frequencies. With 100~MHz channel bandwidth, distance estimation accuracy in the order of 1~m and target detection probability exceeding 90\% were shown to be feasible at SNRs lower than $-30$~dB.

Then, the fundamental TX--RX isolation challenge was addressed, with specific emphasis on monostatic shared-antenna base-station deployments. 
To prevent receiver saturation and to reduce the inherent masking effect of the direct leakage sidelobes, efficient RF and digital cancellation solutions were described, properly tailored to the OFDM radar use case such that frequency-selective TX--RX coupling can be handled while the actual reflections from true targets are preserved. Comprehensive RF measurements were provided, verifying the basic hypothesis as well as concretely show-casing successful sensing of static and moving targets, such as drones and cars, while evidencing measured TX--RX isolation of approximately 100~dB. Our future work will focus on extending the 5G NR based radio sensing research towards the mmWave bands, particularly 28~GHz and 39~GHz that are already standardized for NR. Additionally, we study the potential and feasibility of simultaneously receiving an inband uplink data signal while still sensing the reflections of the downlink transmit signal, all at the same channel.



\bibliographystyle{IEEEtran}
\bibliography{mainReferences}


\begin{IEEEbiography} [{\includegraphics[width=1in,height=1.25in,clip,keepaspectratio]{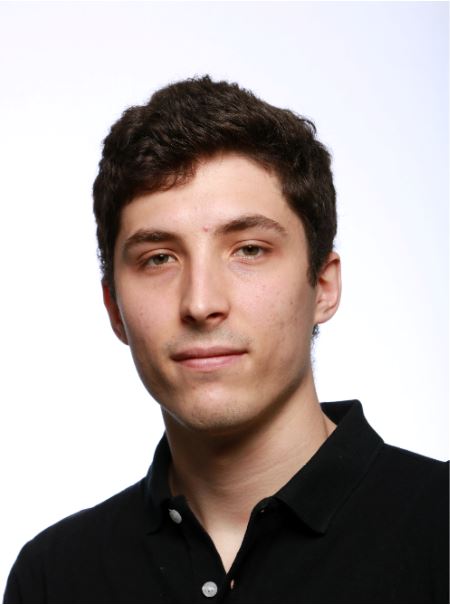}}] {Carlos Baquero Barneto}
(S'18) is a Ph.D. candidate at the Department of Electrical Engineering at Tampere University, Finland. He received his Bachelor and M.Sc. Degrees in Telecommunication engineering from Universidad Politecnica de Madrid (UPM), Madrid, Spain, in 2017 and 2018, respectively. His research interest lies in the area of waveform sharing for radar and communications.
\end{IEEEbiography}

\begin{IEEEbiography}[{\includegraphics[width=1in,height=1.25in,clip,keepaspectratio]{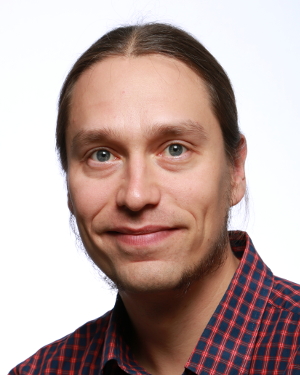}}]{Taneli Riihonen}(S'06--M'14)
received the D.Sc.\ degree in electrical engineering (with distinction) from Aalto University, Helsinki, Finland, in August 2014. He is currently an Assistant Professor (tenure track) at the Faculty of Information Technology and Communication Sciences, Tampere University, Finland. He held various research positions at Aalto University School of Electrical Engineering from September 2005 through December 2017. He was a Visiting Associate Research Scientist and an Adjunct Assistant Professor at Columbia University in the City of New York, USA, from November 2014 through December 2015. He has been nominated eleven times as an Exemplary/Top Reviewer of various IEEE journals and is serving as an Editor for \textsc{IEEE Communications Letters} from October 2014 through January 2019 and for \textsc{IEEE Wireless Communications Letters} since May 2017. He received the Finnish technical sector's award for the best doctoral dissertation of the year in Finland within all engineering sciences and the EURASIP Best PhD Thesis Award 2017. His research activity is focused on physical-layer OFDM(A), multiantenna, relaying and full-duplex wireless techniques with current interest in the evolution of beyond 5G systems.
\end{IEEEbiography}

\begin{IEEEbiography} [{\includegraphics[width=1in,height=1.25in,clip,keepaspectratio]{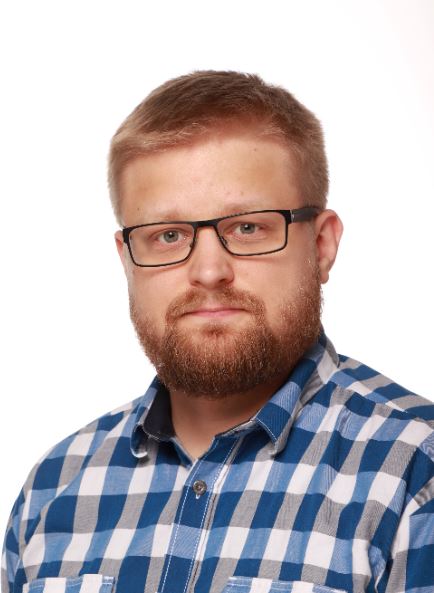}}] {Matias Turunen}
is a research assistant at the Department of Electrical Engineering at Tampere University (TAU), pursuing his M.Sc degree in electrical engineering. His research interests include inband full-duplex radios with an emphasis on analog RF cancellation, OFDM radar, and 5G New Radio systems.
\end{IEEEbiography}

\begin{IEEEbiography}[{\includegraphics[width=1in,height=1.25in,clip,keepaspectratio]{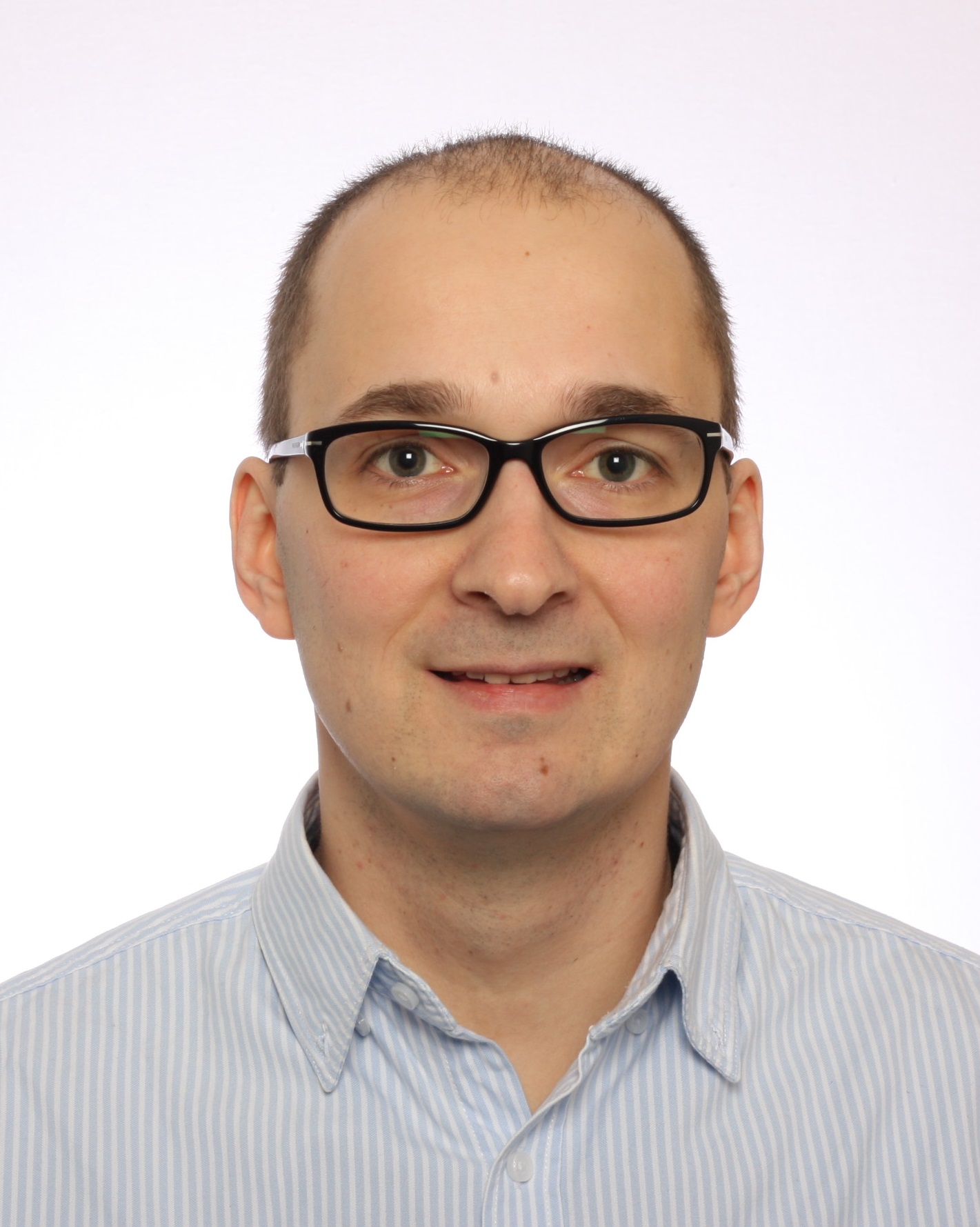}}]{Lauri Anttila}
(S'06--M'11) received the M.Sc. and the D.Sc. (Hons.) degrees in electrical engineering from Tampere University of Technology (TUT), Finland, in 2004 and 2011, respectively. Since 2016, he has been a University Researcher at the Department of Electrical Engineering, Tampere University (formerly TUT). In 2016-2017, he was a Visiting Research Fellow at the Department of Electronics and Nanoengineering, Aalto University, Finland. He has co-authored 100+ refereed articles, as well as three book chapters. His research interests are in radio communications and signal processing, with a focus on the radio implementation challenges in systems such as 5G, full-duplex radio, and large-scale antenna systems.
\end{IEEEbiography}

\begin{IEEEbiography} [{\includegraphics[width=1in,height=1.25in,clip,keepaspectratio]{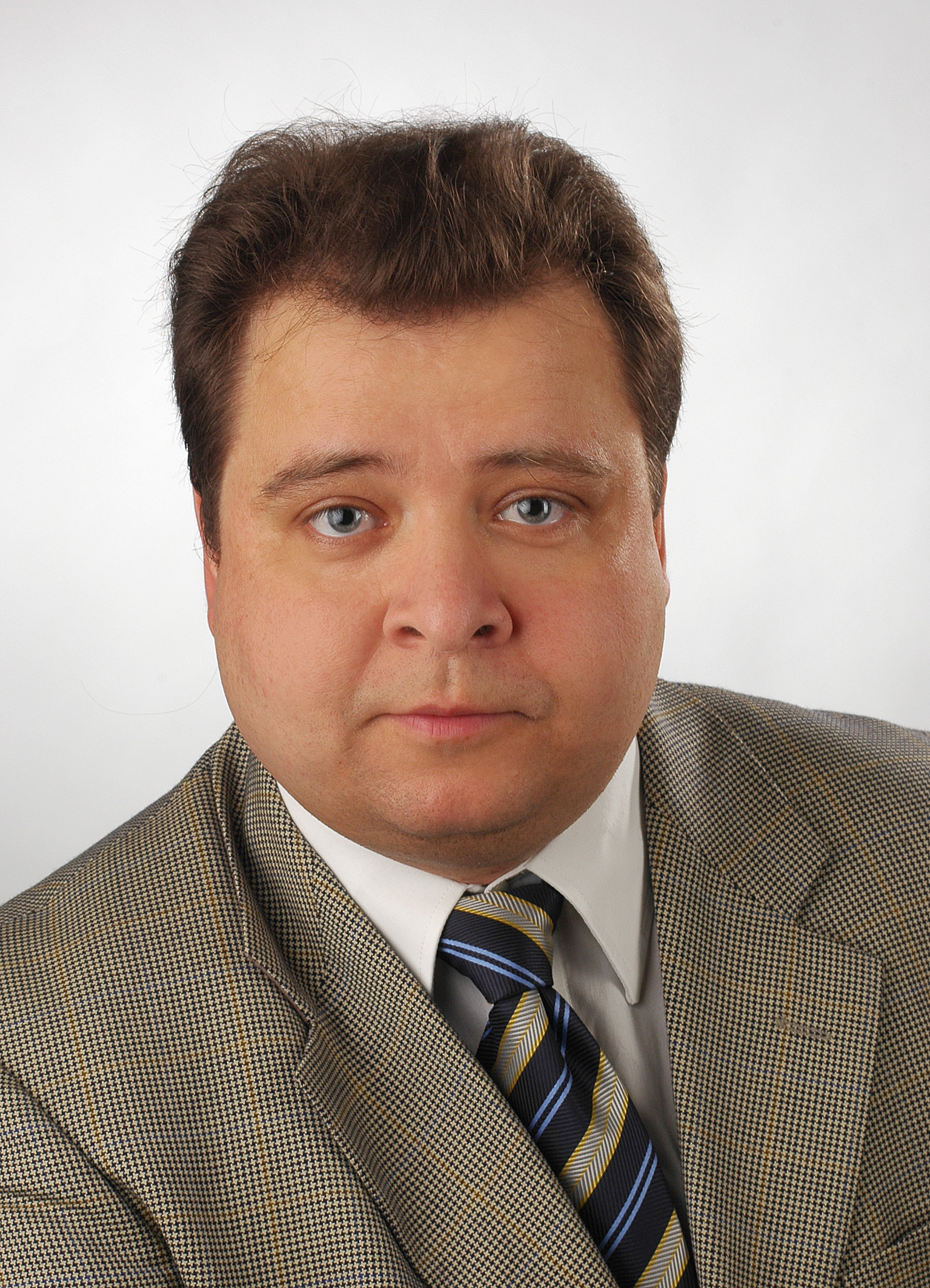}}] {Marko Fleischer}
 received his Diploma Degrees in Electrical Engineering (EE) from TU Dresden, Germany, in 1996. In the late 1990s, he was working in the defense industry and joined Siemens Mobile Communication in 2000. Since than he has been involved in the development of new radio concepts ranging from active to passive intermodulation cancellation methods (AIM/PIM/SIC) , studies of  5G full duplex and OFDM radar. Currently he is working for  Nokia RF System Engineering in Ulm, Germany.
\end{IEEEbiography}

\begin{IEEEbiography} [{\includegraphics[width=1in,height=1.25in,clip,keepaspectratio]{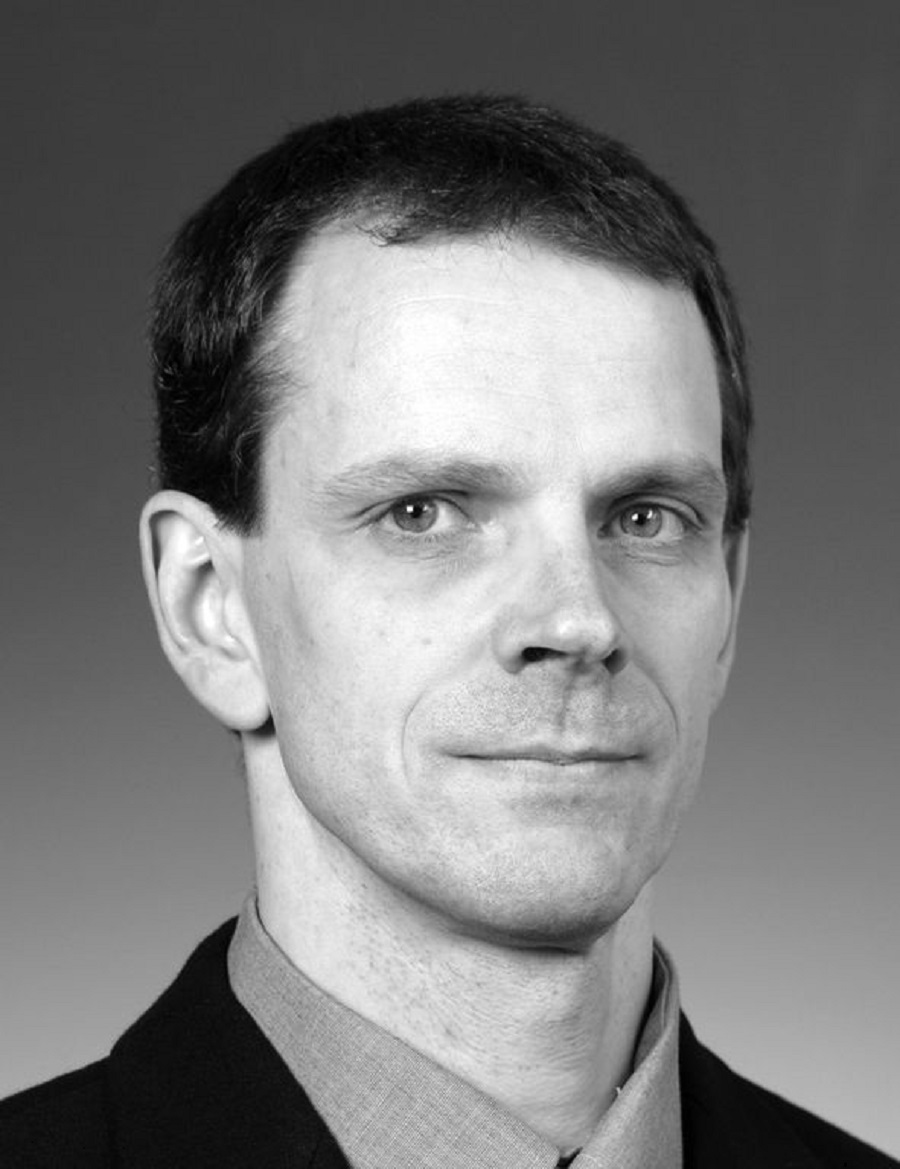}}] {Kari Stadius} (S'95–-M'03) 
received the M.Sc., Lic.Tech., and D.Sc. 
    degrees in electrical engineering from the Helsinki University of 
    Technology, Espoo, Finland, in 1994, 1997, and 2010, respectively. He is 
    currently a Staff Scientist at the Department of Electronics and 
    Nanoengineering, School of Electrical Engineering, Aalto University. He 
    has authored or coauthored over one hundred refereed journal and 
    conference papers in the areas of analog and RF circuit design. His 
    research interests include the design and analysis of RF transceiver 
    blocks with special emphasis on frequency synthesis.
\end{IEEEbiography}

\begin{IEEEbiography} [{\includegraphics[width=1in,height=1.25in,clip,keepaspectratio]{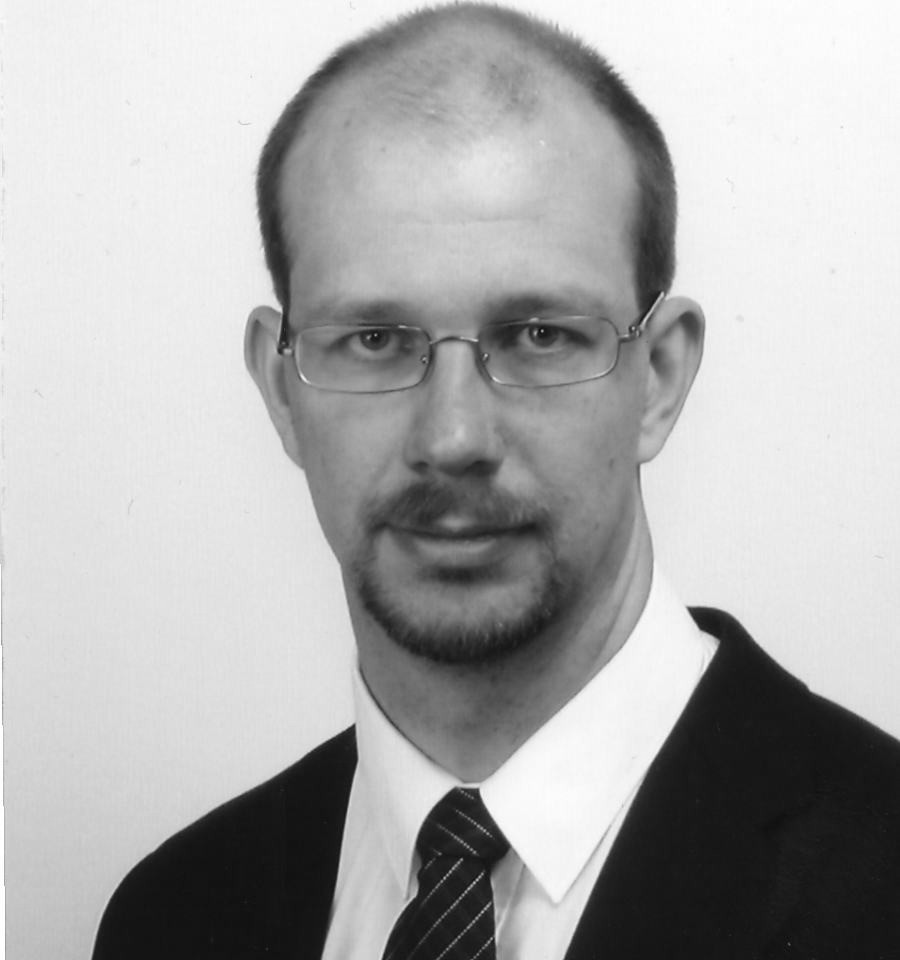}}] {Jussi Ryyn{\"a}nen} (S'99–-M'04–-SM'16) was born in Ilmajoki, Finland, in 1973. 
    He received the M.Sc. and the D.Sc. degrees in electrical engineering 
    from the Helsinki University of Technology, Espoo, Finland, in 1998 and 
    2004, respectively. He is currently a Professor at Aalto University, 
    where he was appointed Head of the Department of Electronics and 
    Nanoengineering in 2016. He has authored or co-authored over 150 
    refereed journal and conference papers in the areas of analog and RF 
    circuit design. He holds seven patents on RF circuits. His main research 
    interests are integrated transceiver circuits for 5G and beyond wireless applications.
\end{IEEEbiography}

\begin{IEEEbiography} [{\includegraphics[width=1in,height=1.25in,clip,keepaspectratio]{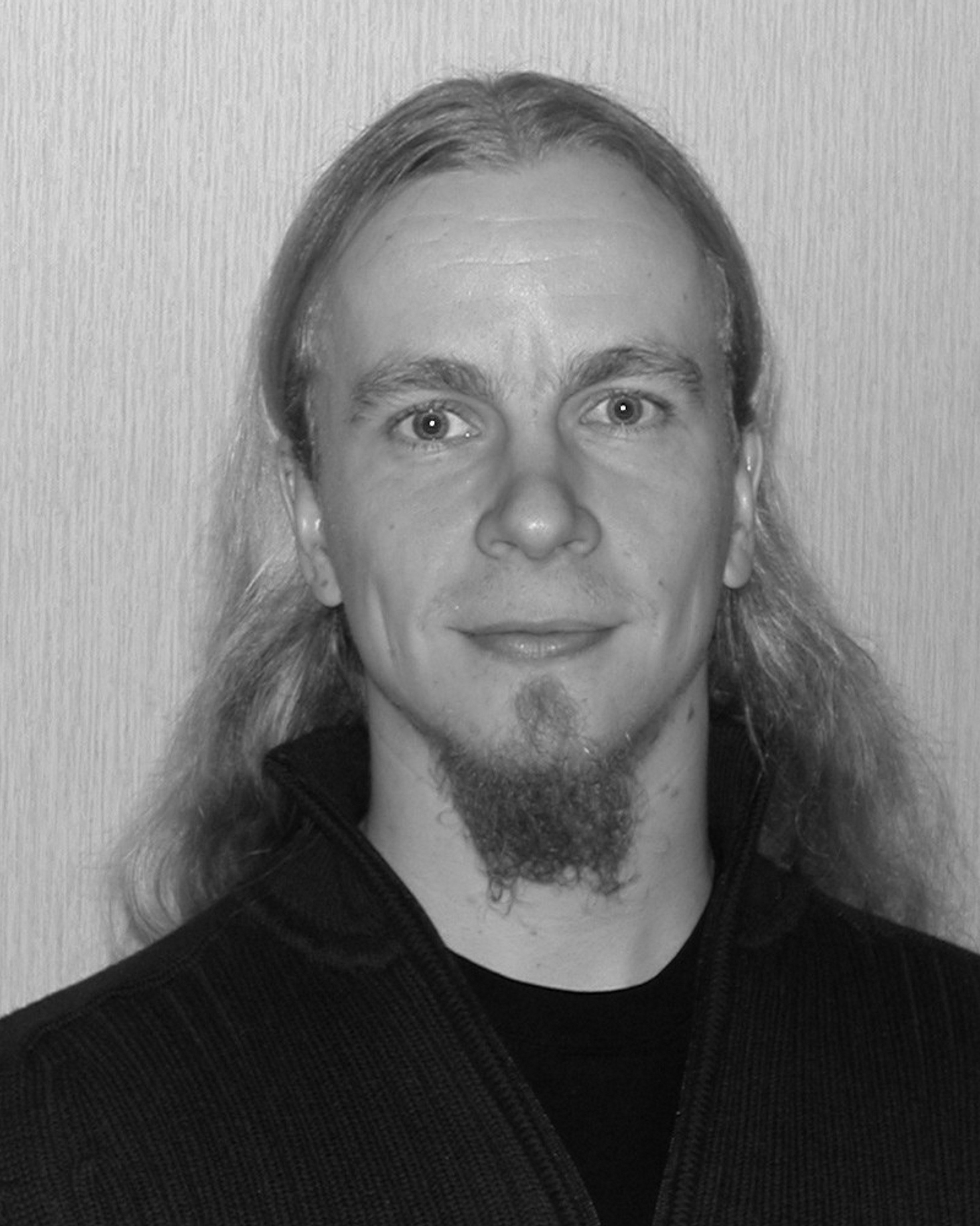}}] {Mikko Valkama}
 received the M.Sc. and D.Sc. Degrees (both with honors) in electrical engineering (EE) from Tampere University of Technology (TUT), Finland, in 2000 and 2001, respectively. In 2002, he received the Best Ph.D. Thesis -award by the Finnish Academy of Science and Letters for his dissertation entitled "Advanced I/Q signal processing for wideband receivers: Models and algorithms". In 2003, he was working as a visiting post-doc research fellow with the Communications Systems and Signal Processing Institute at SDSU, San Diego, CA. Currently, he is a Full Professor and Department Head of Electrical Engineering at the newly formed Tampere University (TAU), Finland. His general research interests include radio communications, radio localization, and radio-based sensing, with particular emphasis on 5G and beyond mobile radio networks.
\end{IEEEbiography}

\end{document}